\documentclass[prb,aps,twocolumn,showpacs,floats]{revtex4}

\usepackage{amssymb}
\usepackage{psfig}
\usepackage{graphicx}
\usepackage{graphics}

\begin{document}

\title{Emergence of quasi-metallic state in disordered 2D electron gas due
to strong interactions }
\author{B. Rosenstein}
\thanks{corresponding author}
\email[E-mail address:]{baruchro@hotmail.com}
\affiliation{National Center for Theoretical Sciences and Electrophysics 
Department,\\
National Chiao Tung University, Hsinchu 30050, Taiwan and \\
Physics
Department, Bar Ilan University, Ramat Gan, Israel.}
\author{\surname{Tran} Minh-Tien}
\affiliation{Electrophysics Department, National Chiao Tung University, 
Hsinchu 30050,
Taiwan and \\
Institute of Physics, National Center for Natural Sciences and Technology,
Hanoi, Vietnam}

\begin{abstract}
The interrelation between disorder and interactions in two dimensional
electron liquid is studied beyond weak coupling perturbation theory. Strong
repulsion significantly reduces the electronic density of states on the
Fermi level. This makes the electron liquid more rigid and strongly
suppresses elastic scattering off impurities. As a result the weak
localization, although ultimately present at zero temperature and infinite
sample size, is unobservable at experimentally accessible temperature at
high enough densities. Therefore practically there exists a well defined
metallic state. We study diffusion of electrons in this state and find that
the diffusion pole is significantly modified due to "mixture" with static
photons similar to the Anderson - Higgs mechanism in superconductivity. As a
result several effects stemming from the long range nature of diffusion like
the Aronov - Altshuler logarithmic corrections to conductivity are less
pronounced.
\end{abstract}

\maketitle





\section{Introduction}

The question of mutual influence of long range Coulomb interactions and
disorder in two dimensional electron gas (2DEG) attracted a great attention
after an unexpected discovery of metallic state and clear metal - insulator
transition by Kravchenko and coworkers. \cite{Kravchenko,Abrahams} The very
existence of a metallic state with finite conductivity at zero temperature
is in conflict with the weak localization theory, \cite{four} which predicts
that in 2D even negligible amount of disorder localizes electrons at
sufficiently low temperature. The theory however was firmly established at
weak coupling or for short range interactions only, while the metallic state
exists and the transition was found for rather strong coupling $%
r_{s}=E_{ee}/E_{F}\sim 10$, where $E_{ee}$ is the average interaction energy
per electron and $E_{F}$ is the Fermi energy. Therefore Coulomb interactions
dominate the kinetic energy and cannot be considered "small". In addition to
an obvious difficulty to treat quantitatively or even qualitatively the
strong coupling, it is not clear which one, disorder or Coulomb
interactions, should be considered as a most important cause of the
transition to an insulating state (the corresponding insulating state in
these cases is of "Anderson" or "Mott" type \cite{Kotliar}). Most probably
it results from a nontrivial combination of these interactions.

The standard approach starts with a commonly accepted argument that a long
range Coulomb interaction after \textquotedblright bubble
resummation\textquotedblright\ of the random phase approximation (RPA) type
\cite{Mahan} becomes effectively short range. Therefore one can start the
treatment of disorder after this resummation was performed. Disorder is
treated within a similar approach in which \textquotedblright
rainbow\textquotedblright\ diagrams \cite{Abrikosov} \textquotedblright
ladders and crossed ladders resummation\textquotedblright\ \cite{Wegner}
(or, more systematically, the \textquotedblright steepest
descent\textquotedblright\ approximations \cite{Belitz} in the path
integrals language \cite{Negele}) with interaction being already short
ranged. In this way two kinds of massless modes determining the properties
of the disordered electron gas are identified: diffusons (describing
diffusive nature of the electron motion due to impurities) and Cooperons in
the particle - particle channel. It is the last which lead to weak
localization due to logarithmic infrared (IR) divergences in leading
fluctuation contribution to conductivity \cite{Vollhardt} (diffusons can
also lead to IR divergences at yet higher orders \cite{Belitz4})
\[
\Delta \sigma ^{wl}(T)=\frac{e^{2}}{\pi h}\log [T\tau /\hbar ],
\]%
where $\tau $ is a free system relaxation time.

More sophisticated renormalization group (RG) based methods using
\textquotedblright path integral\textquotedblright\ \cite{Finkelstein} and
\textquotedblright $\sigma $ - models\textquotedblright\ \cite%
{Wegner,Pruisken,Hikami} with Coulomb interactions 
\cite{Pruiskengen,Baranov}%
\ were developed. Considering high order vertex renormalization, it was
found that there are additional logarithmic IR (Aronov - Altshuler \cite%
{Aronov}) divergencies:%
\[
\Delta \sigma ^{ee}(T)=\frac{e^{2}}{\pi h}(1-3/4F^{\ast })\log [T\tau /\hbar
],
\]%
where $F^{\ast }$ the Fermi surface average of the screened Coulomb
interaction, leading to a conclusion that long interactions increase
tendency to weak localization. \cite{Altshuler} This leads to a difficulty
in understanding recent experiments in which apparently interactions do not
necessarily lead to rapid increase of resistivity. Recent detailed
experimental studies \cite{Coleridge,Rahimi} clearly show that near the
putative metal - insulator transition logarithmic terms either are
suppressed or cancel each other (several arguments were put forward in 
Ref.~%
\onlinecite{Coleridge} against such a fortuitous cancellation). The
conductivity dependence on temperature follows the Gold - Dolgopolov's 
\cite%
{Gold} linear decrease, which at higher temperatures crosses over to the
ballistic regime studied in detail recently in Ref.~\onlinecite{Aleiner}.
Generally within this approach the Coulomb interaction is screened first and
the disorder effects are treated later. However recent electron spin
resonance experiment \cite{Wilamowski} demonstrated that the screening
length rapidly diverges when density is reduced towards the transition
point. The density of states (DOS) at the Fermi level vanishes. It was
noticed long time ago \cite{Giuliani} that the diffusive motion of electrons
slows down the process of screening. Therefore in the limit of small density
and when which disorder seems to play an important and possibly crucial
role, it is reasonable to start from an approximation in which the
interaction is not rendered short range.

With these experimental facts in mind, we reconsider the question of the
inter-relation of disorder and Coulomb interactions in 2DEG within a single
consistent systematic approach without replacing it by a short range
potential from the beginning as is done in the clean limit or high density.
The necessarily nonperturbative approach consists of two steps. First is a
variational one (nonperturbative in coupling) and is similar in spirit the
Hartree - Fock for clean metals or the BCS approximation in superconducting
metals. We find in section II the \textquotedblright best\textquotedblright\
quadratic Hamiltonian representing the system. On this set of Hamiltonians a
quasiparticle (and quasi - hole) Green function is a variational parameter.
There are possible contributions in the particle - particle (Cooper pairs)
as well as the particle - hole channels due to Coulomb interactions (Hartree
state in direct and Fock state in exchange channel), while interactions with
disorder can be treated in a similar manner with the frequency dependent
relaxation time being one of the variational parameters. Possible
condensates in several channels do not realize: of course there is no
condensate in the Coopper channel for a repulsive interaction and there is
also no condensation in the direct channel due to the charge neutrality as
is shown in section IIA. However the strong long range exchange interaction
creates (even in the clean case \cite{Jang}) a dip in the DOS on the Fermi
surface. At infinitely strong coupling the DOS on the Fermi surface
approaches zero (to avoid confusion, this reduction is not related to the
one found at higher orders for screened interaction in Ref.~%
\onlinecite{Altshuler}, see discussion of this topic in section IVD). This
makes the electron liquid very rigid and, as a result, the effects of
disorder are greatly suppressed. This in turn leads to increase in
conductivity at large coupling. The emergence of the above phenomenon can
already be seen on perturbative level. The first order quasiparticle energy
shift due to exchange is:%
\[
\Sigma _{p}=\left\langle p|H_{int}|p\right\rangle =-\sum\limits_{p^{\prime
}}v(p-p^{\prime })\mathrm{sign}\left( \mu -\frac{p^{\prime 2}}{2m^{\ast }}%
\right) .
\]%
It is easily shown (section IIB) that for purely repulsive $v(p)$ energy of
states above the Fermi level is shifted up, while energy of states below the
Fermi level is shifted down. The logarithmic vanishing of the DOS is a
direct consequence of the long range nature of the Coulomb interaction. It
is important to note that the significant reduction of the DOS near the
Fermi level does not mean that the effective mass is smaller than the band
effective mass. On the contrary, it was shown in the clean case \cite%
{TK,Jang} that despite this the effective mass grows with coupling as was
observed recently in Shubnikov - de Haas experiments. \cite{Pudalov3} We
comment more on that in section IIB.

After the variational quadratic Hamiltonian (or variational quadratic action
in the path integral formalism) is found, we introduce in section IIIA a
systematic perturbation theory around it. In the path integral language 
\cite%
{Negele,Belitz} it is a conventional \textquotedblright steepest
descent\textquotedblright\ expansion with the variational action as a saddle
point. First we introduce fields describing various possible kinds of
fluctuations: diffusons, Cooperons, static photons (corresponding to the
direct Coulomb interaction channel), the exchange and the Cooper channel
interactions. The last two are evidently massive as well as half of
diffusons and Cooperons. \cite{Belitz} However we find that there is a
nontrivial mixing between photons and diffusons. The phenomenon is very
reminiscent of the Anderson - Higgs mechanism in superconductivity \cite%
{Anderson} in which massless Goldstone boson of phase is mixing with
(dynamical) photon. As a result both modes become \textquotedblright
massive\textquotedblright . In the case of strongly coupled 2DEG the modes
are not really massive, the density - density correlator describing diffuson
becoming "harder":%
\begin{equation}
\frac{1}{\omega \tau _{r}+e^{2}D_{r}|p|/4\pi },  \label{difprop}
\end{equation}%
compared to the noninteracting diffusion pole%
\begin{equation}
\frac{1}{\omega \tau _{r}+D_{r}p^{2}}.  \label{pole}
\end{equation}%
The (static) photon becomes RPA screened:%
\[
\frac{1}{2|p|/e^{2}+2\pi }.
\]%
The renormalized diffusion constant $D_{r}$ increases (nonperturbatively)
with coupling from its noninteracting value of $D=\mu \tau /m$. Similarly
the renormalized relaxation time $\tau _{r}$ increases with $r_{s}.$

Equation~(\ref{difprop}) implies that electrons at large distances are no
longer \textquotedblright diffusive\textquotedblright : they obey diffusion
equation with first space derivative only. The Cooperon on the other hand
still retains its typical \textquotedblright diffusive\textquotedblright\
pole form Eq.~(\ref{pole}). The approximation scheme at higher orders
therefore nontrivially combines the RPA and the disorder resummation on the
same footing. It is important to emphasize that the scheme is manifestly
\textquotedblright gauge invariant\textquotedblright . As was shown in 
Refs.~%
\onlinecite{Baranov,Pruiskengen}, that it is very important to ensure gauge
invariance at each stage in order not to miss important \textquotedblright
vertex corrections\textquotedblright\ necessary to ensure charge
conservation at each order of the expansion including the variational stage.

Next in section IV we turn to the \textquotedblright fluctuation
corrections\textquotedblright\ leading to weak localization. The leading
fluctuation correction to conductivity is infrared divergent. However, as
the coupling $r_{s}$ grows, the correction grows slower than the main
(Drude) contribution. This justifies the expansion even for small values of 
$%
D<1$ at which the standard $1/D$ expansion is invalid ($D_{r}\gg 1$ is
however required). Therefore the crossover temperature at which conductivity
starts approaching zero is significantly lower than that for non-interacting
electrons. This temperature is estimated as a temperature at which the
perturbation theory in fluctuations breaks down, namely when the correction
becomes a significant fraction of the leading order contribution. The
crossover temperature, according to our analysis becomes unobservably small
since it vanishes exponentially fast with coupling (due to logarithmic
dependence of the fluctuation correction on temperature serving as an IR
cutoff). Therefore one can practically (for samples of finite albeit large
size) talk about stable metallic state in 2DEG. Our conclusions and
discussion of the phase diagram, as well as relations with other approaches
are subject of the concluding section V.

\section{Variational principle for Coulomb interactions in the presence of
disorder}

\subsection{Model and the basic approximation}

\subsubsection{The model}

We consider a system of electrons with effective band mass $m^{\ast }$
confined to a plane interacting with each other and with random potential $%
U(x)$:%
\begin{eqnarray}
H &=&\int_{x}c_{x\sigma }^{\dagger }\left( -\frac{\nabla ^{2}}{2m^{\ast }}%
-\mu +U(x)\right) c_{x\sigma }  \nonumber \\
&&+\frac{1}{2}\int_{x,y}c_{x\sigma _{1}}^{\dagger }c_{x\sigma
_{1}}V(x-y)c_{x\sigma _{2}}^{\dagger }c_{x\sigma _{2}}.  \label{H}
\end{eqnarray}%
We set $\hbar =1$ throughout theoretical parts of the paper. $\sigma $ is a
spin and valley index and $v(x)$ is the 3D Coulomb interaction which has the
following Fourier transform
\begin{equation}
v(p)=\frac{e^{2}}{2\epsilon }\frac{1}{p}\left[ 1-\delta (p)\right] .
\label{v}
\end{equation}%
Here $\epsilon $ the dielectric constant and the last term describes the
background ensuring charge neutrality of the system, $v(p=0)=0$. Passing to
the standard imaginary time path integral formulation \cite{Negele} and
performing a well known replica trick \cite{Belitz} one obtains the action:
\begin{eqnarray}
A[\psi ,\overline{\psi }] &=&\int_{x,t}\overline{\psi }_{xt}^{a\sigma
}\left( \partial _{t}-\frac{\nabla ^{2}}{2m^{\ast }}-\mu \right) \psi
_{xt}^{a\sigma }  \nonumber \\
&&-\frac{1}{2\tau }\int_{x,t,s}\overline{\psi }_{xt}^{a\sigma _{1}}\psi
_{xt}^{a\sigma _{1}}\overline{\psi }_{xs}^{b\sigma _{2}}\psi _{xs}^{b\sigma
_{2}}  \nonumber \\
&&+\frac{1}{2}\int_{x,y,t}\overline{\psi }_{xt}^{a\sigma _{1}}\psi
_{xt}^{a\sigma _{1}}v(x-y)\overline{\psi }_{yt}^{a\sigma _{2}}\psi
_{yt}^{a\sigma _{2}} .  \label{Aspin}
\end{eqnarray}%
Here $\ a,b=1,...,N_{r}$ are replica indices, $\tau $ is the "bare"
relaxation time describing strength of the random potential. Let us first
consider, for the sake of simplicity, the spin polarized case and only one
"valley" (returning to the general case in section IVA), which means that we
drop the spin indices $\sigma $. The path integral formulation of the
variational principle, being completely equivalent to the standard methods
like summation of diagrams or Bogoliubov transformations, allows, in
addition, a convenient treatment of quantum and thermal fluctuations.
Transforming to the Matsubara frequency $\omega _{n}=(2n+1)\pi T$ and the
momentum basis for the Grassmannian fields
\[
\psi _{xt}^{a}=\sqrt{T}\sum\limits_{p}\sum\limits_{n}\exp [i(px-\omega
_{n}t)]\psi _{pn}^{a},
\]%
and separating regions of phase space in which interaction connects
electrons near the Fermi surface, one obtains \cite{Belitz}:%
\begin{equation}
A=\sum\limits_{p}\sum\limits_{n}\overline{\psi }_{pn}^{a}\left( -i\omega
_{n}+\frac{p^{2}}{2m^{\ast }}-\mu \right) \psi _{pn}^{a}+A_{dis}+A_{C} ,
\end{equation}%
\begin{eqnarray*}
A_{dis}=-\frac{1}{2\tau }\sum\limits_{p,q,r}\sum\limits_{n,m}\!\!\!\! &&[%
\overline{\psi }_{p-q,n}^{a}\psi _{q,n}^{a}\overline{\psi }_{-p-r,m}^{b}\psi
_{r,m}^{b} \\
&&+\overline{\psi }_{p-q,n}^{a}\psi _{-p-r,n}^{a}\overline{\psi }%
_{r,m}^{b}\psi _{q,m}^{b} \\
&&+\overline{\psi }_{p-q,n}^{a}\psi _{-p-r,n}^{a}\overline{\psi }%
_{q,m}^{b}\psi _{r,m}^{b}] , \\
A_{C}=\frac{T}{2}\sum\limits_{p,q,r}\sum\limits_{n,m}\!\!\!\! &&[\overline{%
\psi }_{p-q,n}^{a}\psi _{q,n}^{a}v(p)\overline{\psi }_{-p-r,m}^{b}\psi
_{r,m}^{b} \\
&&+\overline{\psi }_{p-q,n}^{a}\psi _{-p-r,n}^{a}v(p)\overline{\psi }%
_{r,m}^{b}\psi _{q,m}^{b} \\
&&+\overline{\psi }_{p-q,n}^{a}\psi _{-p-r,n}^{a}v(p)\overline{\psi }%
_{q,m}^{b}\psi _{r,m}^{b}] .
\end{eqnarray*}%
All the fermion's momenta $p,q,r$ are now considered to be around $p_{F}$.
Both the disorder and the Coulomb interaction parts, $A_{dis}$ and $A_{C}$
respectively, have three terms corresponding to direct (Hartree), exchange
(Fock) electron - hole channels and the electron - electron (Cooper) 
channel.

\subsubsection{The most general quadratic Hamiltonian and the Hubbard -
Stratonovich fields}

A convenient way to look for the most general quadratic action is to perform
a Hubbard-Stratonovich (HS) transformation introducing a field for each of
the six channels. We should not consider the direct channel for disorder
though since it is of higher order in number of replicas $N_{r}$ which
should approach zero (we assume that replica symmetry is not broken
spontaneously). The effective action in terms of these fields is rather
complicated:
\begin{eqnarray}
\!\!\! &A_{eff}&\!\!\!\!=-\mathrm{Tr} \mathrm{Log} \left[ G_{N}^{-1}\right]
+\sum\limits_{pnm}[Q_{pnm}^{ab}Q_{pnm}^{\ast ab}+\Delta _{pnm}^{\ast
ab}\Delta _{pnm}^{ab}]  \nonumber \\
&+\frac{1}{2}\sum\limits_{pp^{\prime }n}&\!\![\Phi _{pn}^{\ast
}v^{-1}(p-p^{\prime })\Phi _{p^{\prime }n}+\Theta _{pp^{\prime }n}^{\ast
a}v^{-1}(p-p^{\prime })\Theta _{pp^{\prime }n}^{a}  \nonumber \\
&&+E_{pp^{\prime }n}^{\ast ab}v^{-1}(p-p^{\prime })E_{pp^{\prime }n}^{ab}\ ]
.  \label{Aeff}
\end{eqnarray}%
Here $G_{N}$ is fermionic Green's function in Nambu space defined by
\begin{equation}
\eta _{pn}^{a}=\frac{1}{\sqrt{2}}\left(
\begin{array}{c}
\psi _{pn}^{a} \\
-\overline{\psi }_{-pn}^{a}%
\end{array}%
\right) ;\overline{\eta }_{pn}^{a}=\frac{1}{\sqrt{2}}\left(
\begin{array}{cc}
\overline{\psi }_{pn}^{a} & -\psi _{-pn}^{a}%
\end{array}%
\right) .  \label{Nambueta}
\end{equation}%
The inverse fermionic propagator is a $2\times 2$ matrix \ \
\begin{equation}
G_{N}^{-1}=\left(
\begin{array}{cc}
G^{-1} & F \\
F^{\ast } & G^{-1}{}^{\ast }%
\end{array}%
\right) ,  \label{Nambu}
\end{equation}%
whose elements $G^{-1}$ and $F$ are themselves matrices in $p,n$ and $a$
space. The diagonal element is:
\begin{eqnarray}
\left\langle _{p^{\prime }n^{\prime }}^{a}|G^{-1}|_{pn}^{b}\right\rangle
&=&\delta _{nn^{\prime }}\delta ^{ab}\left[ \left( i\omega 
_{n}-\frac{p^{2}}{%
2m^{\ast }}+\mu \right) \delta _{pp^{\prime }}+i\Phi _{pn}\right]  \nonumber
\\
&&+\frac{i}{\sqrt{\tau }}Q_{p-p^{\prime }nn^{\prime }}^{ab}-\delta
_{nn^{\prime }}E_{pp^{\prime }n}^{ab},  \label{Gm}
\end{eqnarray}%
where the field $Q$ describes the diffuson, $\Phi $ is a static photon field
in the direct channel, while $E$ is an \textquotedblright exchange
field\textquotedblright . The off - diagonal element
\begin{equation}
\left\langle _{p^{\prime }n^{\prime }}^{a}|F|_{pn}^{b}\right\rangle 
=\frac{i%
}{\sqrt{\tau }}\Delta _{pp^{\prime }nn^{\prime }}^{ab}+\delta ^{ab}\Theta
_{pp^{\prime },n+n^{\prime }}^{a}  \label{F}
\end{equation}%
contains the $\Delta $ and the $\Theta $ fields describing the Cooperon
channel in the disorder part and the Cooper channel in the Coulomb
interaction part (if the interaction were attractive this channel would have
lead to superconductivity) respectively. This effective action should be
minimized as a function of all five HS fields determining the fermionic
Green function. The solution of this variational problem is discussed in the
following subsection. Later in section III we expand the path integral
around the solution of the minimization equations to quadratic order
(harmonic approximation) to determine elementary excitation modes and then
in section IV use Feynman rules to compute fluctuation corrections.

\subsubsection{The saddle point equations}

Minimization of the effective action Eq.~(\ref{Aeff}) poses a nontrivial
mathematical problem. Let us first remove obviously irrelevant fields and
functional dependencies using symmetry arguments. Since the Coulomb
interaction is purely repulsive, we assume that the electromagnetic $U(1)$
gauge symmetry is unbroken (there is no condensation of the electron -
electron pairs). Therefore $\Delta _{SP}=\Theta _{SP}=0$. This by no means
indicates that there are no fluctuations in these channel. On the contrary
fluctuations in the Cooperon channel play an important role in destroying
the metallic state.

The translation invariance in space and time (we assume that the ground
state is a liquid rather than Wigner crystal \cite{Chakravarty}) and the
unbroken replica symmetry (assuming the ground state is a disordered,
possibly overcooled liquid rather then electron glass \cite{Dobrosavljevic})
implies:%
\begin{eqnarray}
Q_{pnmSP}^{ab} &=&\delta ^{ab}\delta _{nm}\delta _{p}q_{n},  \nonumber \\
E_{pp^{\prime }nSP}^{ab} &=&\delta ^{ab}\delta _{n}\delta _{pp^{\prime
}}e_{p},  \label{assumptions} \\
\Phi _{pn} &=&\delta ^{ab}\delta _{n}\delta _{p}\phi .  \nonumber
\end{eqnarray}%
We will comment on the last two (nontrivial) assumptions in section V.
Consequently the inverse Green's function of fermions simplifies to%
\begin{eqnarray}
F &=&0,  \nonumber \\
\left\langle _{p^{\prime }n^{\prime }}^{a}|G^{-1}|_{pn}^{b}\right\rangle
&=&\delta ^{ab}\delta ^{nn^{\prime }}\delta (p-p^{\prime })(G_{p}^{n})^{-1},
\label{Gm1} \\
(G_{p}^{n})^{-1} &=&i\left( \omega _{n}+\frac{q_{n}}{\sqrt{\tau }}\right)
+\mu -\frac{p^{2}}{2m^{\ast }}-\Sigma _{p}.  \nonumber
\end{eqnarray}%
The minimization equation for the static photon condensate $\phi $ is
\begin{equation}
\phi =-i\sqrt{T}\delta (p)v(p)\sum\limits_{q,n}G_{q}^{n}.  \label{SPfi}
\end{equation}%
However due do neutralizing background $v(p=0)=0,$ and the right hand side
of this equation vanishes. Therefore $\phi =0$. Minimization equation for $%
q_{n},e_{p}$ are
\begin{eqnarray}
q_{n} &=&\frac{i}{\sqrt{\tau }}\sum\limits_{q}G_{q}^{n},  \label{SPq} \\
\Sigma _{p} &=&-T\sum\limits_{p^{\prime },n}v(p-p^{\prime })G_{p^{\prime
}}^{n}.  \label{SPe}
\end{eqnarray}%
We start with exact solution of these equations in the clean case.

\subsection{The clean limit}

\subsubsection{A major simplification in the clean limit}

\begin{figure}[t]
\centerline{ \psfig{file=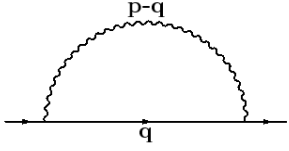,width=0.25\textwidth}}
\caption{The self-energy corresponding to the saddle point equation in the
clean limit}
\label{fig1}
\end{figure}

In the absence of disorder $\tau \rightarrow \infty $, $q\rightarrow 0$ and
we should consider the second equation (\ref{SPe}) only. Substituting the
Green function Eq.~(\ref{Gm1}), it takes a form%
\begin{equation}
\Sigma _{p}=-T\sum\limits_{q,n}\frac{v(\mathbf{p}-\mathbf{q})}{i\omega
_{n}+\mu -q^{2}/2m^{\ast }-\Sigma _{q}},  \label{SPclean}
\end{equation}%
corresponding to the diagram on Fig.~\ref{fig1}. At zero temperature (to
which we confine ourselves for most of the paper) the summation over
Matsubara frequencies results in
\begin{eqnarray}
\Sigma _{p} 
&=&-\frac{1}{2}\sum\limits_{q}v(|\mathbf{p}-\mathbf{q}|)\mathrm{%
sign}[\mu -q^{2}/2m^{\ast }-\Sigma _{q}]  \nonumber \\
&=&-\frac{1}{2\left( 2\pi \right) ^{2}}\int\limits_{q=0}^{\infty
}\int\limits_{\varphi =0}^{\pi }qv\left( \sqrt{p^{2}+q^{2}-2pq\cos (\varphi 
)%
}\right)  \nonumber \\
&&\mathrm{sign}[\mu -q^{2}/2m^{\ast }-\Sigma _{q}],  \label{SPclean1}
\end{eqnarray}%
where $\varphi $ is an angle between fermion's momenta vectors $\mathbf{p}$
and $\mathbf{q}$ and $p,$ $q$ are their lengths. Integral over the angle
gives
\[
\Sigma _{p}=-\frac{e^{2}}{\left( 2\pi \right) ^{2}}\int\limits_{|q|\geq 0}%
\frac{qK[-\frac{4pq}{(p-q)^{2}}]}{|p-q|}\mathrm{sign}[\mu -q^{2}/2m^{\ast
}-\Sigma _{q}].
\]%
Here $K[x]$ is the full elliptic integral of the first kind. 
\cite{Gradstein}
The interaction with neutralizing background amounts to subtracting the term
$q=p$. Now we switch to dimensionless variables describing deviations of
particle's energy from Fermi surface%
\begin{equation}
p\cong \sqrt{2m^{\ast }\mu }(1+\varepsilon +...);\text{ \ \ }q\cong \sqrt{%
2m^{\ast }\mu }(1+\varepsilon ^{\prime }+...),  \label{rescale}
\end{equation}%
rescaling also the variational self energy function $\Sigma _{p}=2\mu
e_{\varepsilon }$. In the resulting integral equation%
\begin{eqnarray}
e_{\varepsilon } &=&\frac{r_{s}}{\sqrt{2}\pi }\int_{\varepsilon ^{\prime
}}\kappa \lbrack \varepsilon -\varepsilon ^{\prime }]\mathrm{sign}%
[\varepsilon ^{\prime }+e_{\varepsilon ^{\prime }}],  \label{SPclean4} \\
\kappa \lbrack \varepsilon ] &\equiv &\frac{K[-4/\varepsilon ^{2}]}{%
|\varepsilon |},  \nonumber
\end{eqnarray}%
we extend the integration over $\varepsilon ^{\prime }$ to $\{-\infty
,\infty \}$ and observe that there exists a solution obeying physically
reasonable property that the sign of $e_{\varepsilon }$ is the same as that
of $\varepsilon $ (we checked that there are no other solutions in
disordered case as well). This makes the right hand side independent of $%
e_{\varepsilon }$ and we obtain a solution%
\begin{equation}
e_{\varepsilon }=\frac{\sqrt{2}r_{s}}{\pi }\int_{\varepsilon ^{\prime
}=0}^{\varepsilon }\kappa \lbrack \varepsilon ^{\prime }]\equiv 
\frac{\sqrt{2%
}r_{s}}{\pi }\kappa _{1}[\varepsilon ],  \label{SPclean5}
\end{equation}%
where function $\kappa _{1}[\varepsilon ]$ can be expressed via Meijer
function, $G_{pq}^{mn}\left( z\Big|\null_{b_{1},\dots ,b_{q}}^{a_{1},\dots
,a_{p}}\right) $, \cite{Gradstein}%
\begin{eqnarray}
\kappa _{1}[\varepsilon ] &=&\frac{1}{4}\mathrm{sign}[\varepsilon
][M(\varepsilon )-M(0)];  \label{M} \\
M(\varepsilon ) &=&G_{33}^{22}\left( \frac{4}{\varepsilon ^{2}}\Big|\null%
_{0,0}^{1/2,1/2,1}\right) .  \nonumber
\end{eqnarray}%
It behaves as $\varepsilon (\log [8/\varepsilon ]+1)/2$ at
small $\varepsilon $ and as $\pi\log [2\varepsilon ]/2$ at large $%
\varepsilon $. The standard dimensionless coupling
\[
r_{s}\equiv \frac{e^{2}}{4\pi \epsilon }\sqrt{\frac{m^{\ast }}{\mu }}
\]%
was introduced.

\begin{figure}[t]
\centerline{ \psfig{file=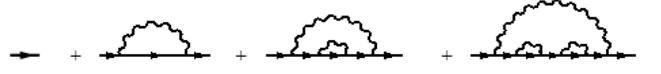,width=0.5\textwidth}}
\caption{The "rainbow" diagrams corresponding to the saddle point equation.
The solid lines denote the fermionic propagator, the wavy lines denote the
Coulomb interactions.}
\label{fig2}
\end{figure}
Generally the solution of the saddle point equation corresponds in terms of
diagrams to summation of all the photonic \textquotedblright
rainbows\textquotedblright , see Fig.~\ref{fig2}. However, as it is well
known, \cite{Mahan} the sum of the diagrams except that of Fig.~\ref{fig1}
vanishes identically. This will no longer be the case in the disordered
electron gas. The fermionic Green function near the Fermi surface, namely at
small $\varepsilon $ becomes nonanalytic%
\begin{equation}
G_{\varepsilon }^{-1}\sim i\widehat{\omega} _{n}+\varepsilon (1+\frac{r_{s}}{\sqrt{2}%
\pi }\log \frac{8}{\varepsilon })  \label{Gmapp}
\end{equation}%
and it is dominated by the interaction "corrections". Here 
$\widehat{\omega}_n  = \omega_n/2 \mu$ is dimensionless frequency.

\subsubsection{Depletion around the Fermi level}

Although no energy gap (the Coulomb gap) have been opened within this
approximation, it follows directly from Eq.~(\ref{Gmapp}) that the DOS right
on the Fermi level vanishes logarithmically:%
\begin{equation}
N(\varepsilon )\sim \frac{1}{\log (1/\varepsilon )}.  \label{DOSclean}
\end{equation}%
Therefore one can term electron gas with such properties \textquotedblright
very marginal Fermi liquid\textquotedblright . The logarithmic dip in the
DOS is even weaker than a power normally associated with such a situation.
\cite{Varma} As we will see in the next subsection, this will naturally lead
to effective reduction of scattering off impurities thereby increasing
conductivity and making the transition to Anderson insulator at least more
difficult. Of course this result is \textquotedblright
perturbative\textquotedblright\ in a sense that for the inverse propagator
only one diagram was taken, Fig.~\ref{fig1}. Therefore the effect
\textquotedblright starts\textquotedblright\ and can be understood at weak
coupling. One can interpret the minimization equations (or the Hartree -
Fock resummation) as a renormalization of energies of the one particle
states due to the collective effect of the many - body electron - electron
repulsions. As we mentioned in Introduction the reduction in DOS in
perturbation theory can be seen from the eigenvalue shifts. Indeed the $%
\mathrm{sign}[\varepsilon ]$ factor in Eq.~(\ref{SPclean5}) makes it clear
that energy of states above the Fermi level are shifted up, while energy of
states below the Fermi level are shifted down. The logarithmic singularity
is a direct consequence of the long range nature of the Coulomb interaction.

Obviously, if there would not be a disorder to interfere with the screening,
the RPA type of reasoning would imply that the singularity would be
\textquotedblright smoothed away\textquotedblright\ or \textquotedblright
cured\textquotedblright\ by the quantum fluctuations corrections. Higher
orders in coupling are increasingly singular and the singularities should be
"resumed away". \cite{Mahan} However even for the screened interactions
there is a dip in the DOS. \cite{Giuliani2} It cannot approach zero, but
might be reduced significantly. We calculated $N(0)$ for different couplings
assuming RPA potential instead of $v(p)$. For several couplings the
reduction of DOS is given in the last line of Table I. It provides an
indication at what degree of disorder we can continue to use the HF approach
without encountering strong screening effects. As one can see, the RPA is
apparently less important to the reduction of DOS already at $r_{s}$ as low
as $1$.

\begin{table}[tbp]
\caption{ The density of states at Fermi level $N(0)$ (normalized to the
ideal gas DOS $N_{0}=1/2\protect\pi $) for various couplings $r_{s}$ and
diffusion constants $\protect\lambda$.}\vspace{.2cm} 
\begin{tabular}{||c|cccccc||}
\hline\hline
{\large {$_\lambda$}} \hspace{-0.5cm} {\Huge {$\diagdown$}} \hspace{-.5cm}%
{\Large {$^{^{r_{s}}}$}} & $0.1$ & $1$ & $2$ & $4$ & $8$ & $16$ \\ \hline
$8$ & $0.873$ & $0.395$ & $0.219$ & $0.106$ & $0.0476$ & $0.0204$ \\
$2$ & $0.892$ & $0.466$ & $0.268$ & $0.129$ & $0.0570$ & $0.0246$ \\
$0.5$ & $0.895$ & $0.557$ & $0.344 $ & $0.167$ & $0.0714$ & $0.0297$ \\
$0.1$ & $0.845$ & $0.651$ & $0.471$ & $0.241$ & $0.103$ & $0.0395$ \\
RPA & $0.961$ & $0.817$ & $0.709$ & $0.623$ & $0.506$ & $0.393$ \\
\hline\hline
\end{tabular}%
%
%
%
%
\end{table}

This is consistent with results of more elaborate calculations of the
renormalization constant $Z$ defined in Eq.~(\ref{Z}) below involving the
Hubbard function in Ref.~\onlinecite{Jang}. The effect of screening is much
smaller at large coupling due to small density of the polarizing electrons.
As we will see in the next subsection, in the disordered case the
\textquotedblright marginality\textquotedblright\ is replaced at large
coupling by large nonperturbative renormalization of the parameters of the
disordered Fermi liquid.

\subsubsection{Effective mass increase vs the DOS drop}

The reduction in the DOS due to the repulsive interaction is related and is
sometimes confused with the issue of the vanishing in Hartree - Fock
approximation of the \textquotedblright renormalized mass\textquotedblright\
and the Fermi liquid parameter $F_{1}^{s}$ defined by \cite{Abrikosov,Mahan}
\begin{equation}
\frac{m_{r}^{\ast }}{m^{\ast }}=\frac{Z}{Z_{F}}=1+F_{1}^{s},  \label{mr}
\end{equation}%
where renormalization constant are defined by%
\begin{eqnarray}
Z^{-1} &=&1+\frac{m^{\ast }}{p_{F}}\frac{\partial }{\partial p}\mathop{\rm
Re}\Sigma ^{ret}(p,\omega =0)\bigg|_{p=p_{F}}  \nonumber \\
&\approx &\frac{1}{2\pi N(\omega =0)},  \label{Z} \\
Z_{F}^{-1} &=&1-\frac{\partial }{\partial \omega }\mathop{\rm Re}\Sigma
^{ret}(p=p_{F},\omega )\bigg|_{\omega =0}.  \nonumber
\end{eqnarray}%
Within the Hartree - Fock approximation (without the RPA resummation) the
retarded self energy $\Sigma ^{ret}(p,\omega )$ does not depend on
frequency. Consequently, $Z_{F}=1$ and since $Z<1$ for repulsive
interactions \cite{Giuliani} the renormalized mass $m_{r}^{\ast }$ is
smaller than the band mass $m^{\ast }$ ($m_{r}^{\ast }$ vanishes for the
long range interactions, see Eq.~(\ref{DOSclean})). Several groups tried to
improve this beyond Hartree - Fock using the RPA and the Hubbard
approximation. \cite{TK,Jang} Generally the wave function renormalization $%
Z_{F}<1$ since it represents departure of the momentum distribution from the
ideal Fermi gas one. Therefore the question whether the effective mass is
larger than the band mass depends on which of the reduction factors spacial 
$%
Z$ or temporary $Z_{F}$ is smaller. Ting, Lee and Quinn \cite{TK} obtained
finite monotonically increasing $m_{r}^{\ast }>m^{\ast }$, while more recent
calculation \cite{Jang} indicates that at $r_{s}<1$, $m_{r}^{\ast }<m^{\ast
} $ and become larger above $r_{s}=1.$ Numerical simulations of the same
system \cite{Kwon} indicate that, on the one hand side the renormalized mass
is definitely smaller than $m^{\ast }$ at least for $r_{s}<5$, but, on the
other hand it increases with $r_{s}$. In recent experiments the renormalized
mass was measured using Shubnikov - de Haas oscillations in magnetic field.
\cite{Pudalov3} Apparently that the renormalized mass deduced that way
monotonically increases with $r_{s}$. To conclude, the increase of the
effective mass with coupling does not necessarily imply that the DOS at
Fermi level cannot drop significantly. This is important for our approach
since the drop in the DOS naively should result in suppression of the
elastic scattering off impurities due to reduced phase space available. Now
we return to the general case of disordered strongly coupled 2DEG.

\subsection{The saddle point equations in the general case}

\subsubsection{Numerical solution}

\begin{figure}[b]
\centerline{ \psfig{file=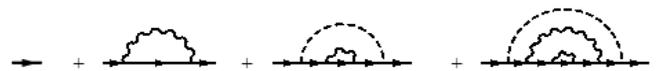,width=0.5\textwidth}}
\caption{The "rainbow" diagrams involving both Coulomb interactions (wavy
lines) and interactions with disorder (dashed lines) in the general case}
\label{fig3}
\end{figure}

Following in the more complicated disordered case 
Eq.~(\ref{SPq})-(\ref{SPe}%
) the same steps as in the clean case, the equations for the scaled
(dimensionless) quantities ($\Sigma _{p}=2\mu e_{\varepsilon },$ $%
q_{n}\equiv 2\mu \sqrt{\tau }\widehat{q}_{\omega },$ $\omega \equiv 2\mu
\widehat{\omega }$) at zero temperature are%
\begin{eqnarray}
\widehat{q}_{\omega } &=&\frac{1}{\left( 2\pi \right) 2\lambda }%
\int_{\varepsilon }\frac{\widehat{\omega }+\widehat{q}_{\omega }}{\left(
\varepsilon +e_{\varepsilon }\right) ^{2}+(\widehat{\omega }+\widehat{q}%
_{\omega })^{2}},  \label{qeq} \\
e_{\varepsilon } &=&\frac{r_{s}}{\sqrt{2}\pi ^{2}}\int_{\omega ,\varepsilon
^{\prime }}\kappa \lbrack \varepsilon -\varepsilon ^{\prime }]\frac{%
\varepsilon ^{\prime }+e_{\varepsilon ^{\prime }}}{\left( \varepsilon
^{\prime }+e_{\varepsilon ^{\prime }}\right) ^{2}+(\widehat{\omega }+%
\widehat{q}_{\omega })^{2}},  \label{eeq}
\end{eqnarray}%
where $\lambda =\mu \tau $  and function
$\kappa \lbrack \varepsilon ]$ was defined in Eq.~(\ref{SPclean4}). The
approximation corresponds to summing up the whole set of "rainbow" diagrams
involving both Coulomb interactions and interactions with disorder, 
Fig.~\ref%
{fig3}.

\begin{figure}[t]
\centerline{ \psfig{file=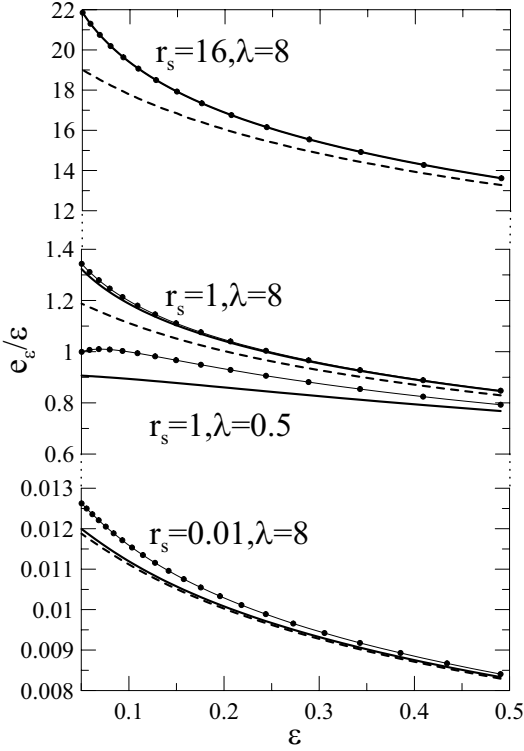,width=0.45\textwidth}}
\caption{The dependence of $e_{\protect\varepsilon }/\protect\varepsilon $
on $\protect\varepsilon $ for various values of $r_{s}$ and $\lambda$. The solid
lines are the numerical results, the dashed lines are the results of $r_{s}$
expansion, and the dots are the results of $1/\protect\lambda$ expansion. }
\label{fig4}
\end{figure}
The results for the real part of self energy for $\lambda =0.5$, $r_{s}=1$
(where in the non-interacting case Anderson localization is very effective)
and $\lambda =8$ , $r_{s}=0.01$, $1$, and $16$ (still a quasi-metal with
small localization effects) are given on Fig.~\ref{fig4}. The corresponding
imaginary part of self energy $\widehat{q}_{\omega }$ is given on Fig.~\ref%
{fig5}. We solved the equations for various values of coupling $r_{s}$
between weak coupling up to $r_{s}=24$ (experimentally the metal - insulator
transition is observed from $r_{s}=15$, \cite{Kravchenko} to $r_{s}=25$ in
GaAs/AlGaAs \cite{Pepper} heterojunctions and around $r_{s}=10$ in Si MOSFET
\cite{Pudalov,Pepper}). Solid lines are simulation results. The dashed lines
are results of expansion in $r_{s}$ briefly described in Appendix, while
dots are the results of the next to leading order in $1/\lambda $ also
summarized in Appendix.

\begin{figure}[t]
\centerline{ \psfig{file=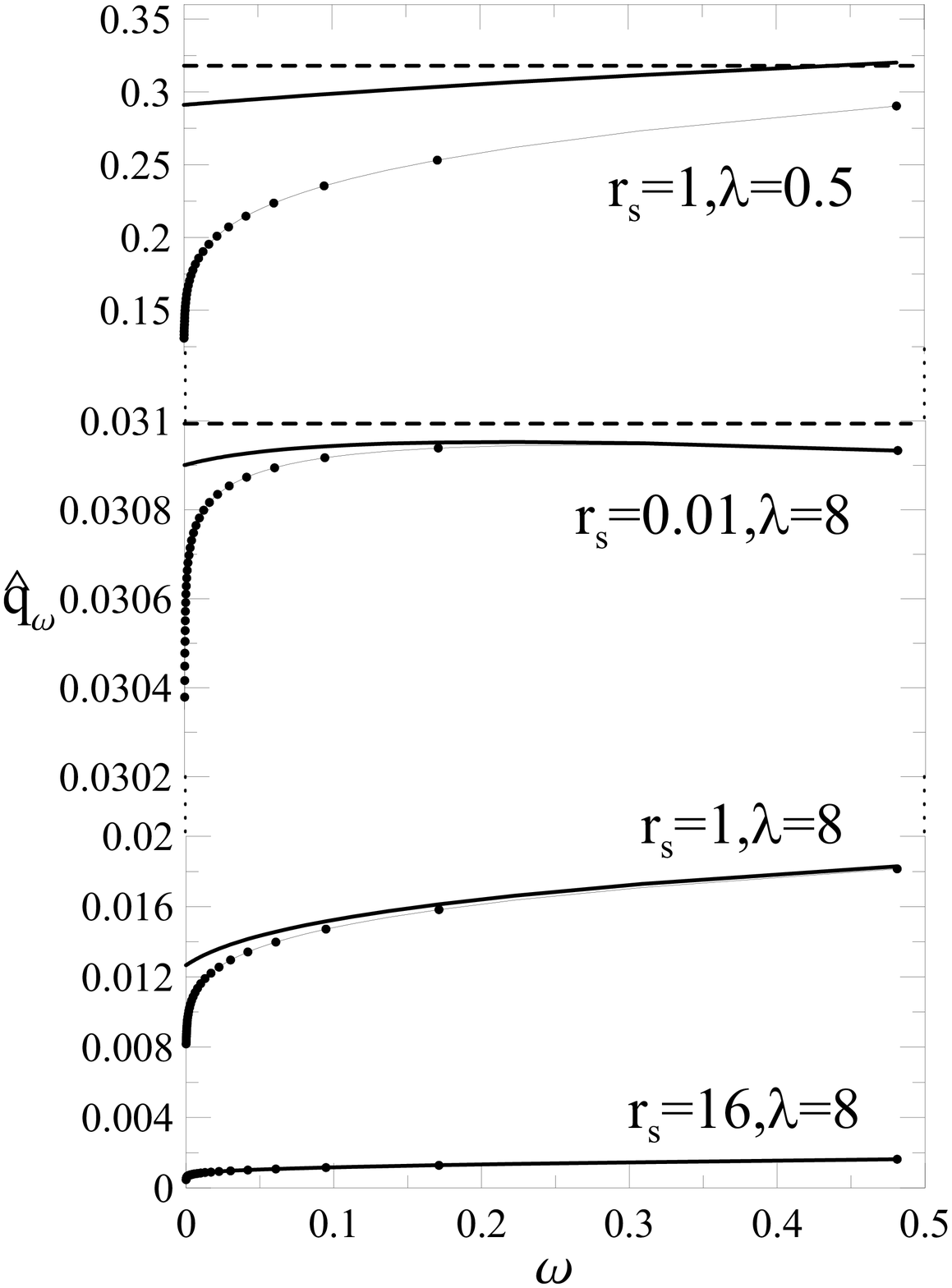,width=0.45\textwidth}}
\caption{The dependence of ${\widehat{q}}_{\protect\omega }$ on $\protect%
\omega $ for various values of $r_{s}$ and $\protect\lambda$. The solid
lines are the numerical results, the dashed lines are the results of $r_{s}$
expansion, and the dots are the results of $1/\protect\lambda$ expansion. }
\label{fig5}
\end{figure}

At zero frequency the large coupling value of $\widehat{q}_{\omega =0}$ is
much smaller than the non-interacting result $1/(4\lambda )$. We define the
renormalized relaxation time via
\begin{equation}
\tau _{r}\equiv \frac{\sqrt{\tau }}{2q_{\omega =0}}=\frac{\tau }{4\lambda
\widehat{q}}.  \label{tau_r}
\end{equation}%
One observes on Fig.~\ref{fig5} that for $r_{s}<10$ the disorder parameter 
$%
\widehat{q}_{\omega }$ is almost independent of $\omega $ in the whole
region where it is important for calculations of integrals over Green's
functions (namely when $\widehat{q}<\omega $). Therefore one can still
consider the system as a disordered liquid without temporary dispersion. It
is a good approximation to simplify the analysis by considering a
variational principle with constant $\widehat{q}_{\omega }=\widehat{q}$.

The saddle point equations in that case take a simpler form (after
integration over $\omega $)%
\begin{eqnarray}
1 &=&\frac{1}{\left( 2\pi \right) 2\lambda }\int_{\varepsilon }\frac{1}{%
\left( \varepsilon +e_{\varepsilon }\right) ^{2}+\widehat{q}^{2}},
\label{simpQ} \\
e_{\varepsilon } &=&\frac{\sqrt{2}r_{s}}{\pi ^{2}}\int_{\varepsilon ^{\prime
}}\kappa \lbrack \varepsilon -\varepsilon ^{\prime }]\mathrm{sign}%
[\varepsilon ^{\prime }]\left( \frac{\pi }{2}-\arctan \frac{\widehat{q}}{%
|\varepsilon ^{\prime }+e_{\varepsilon ^{\prime }}|}\right) .  \nonumber \\
&&  \label{simpE}
\end{eqnarray}%
From Fig.~\ref{fig4} we observe that at very large coupling it is similar to
the \textquotedblright very marginal Fermi liquid\textquotedblright\ of the
clean case.\ The interaction \textquotedblright
correction\textquotedblright\ $\Sigma _{\varepsilon }$ dominates and is not
proportional to $\varepsilon $, therefore it does not reduces to a mere
renormalization of the density of states on the Fermi level. At finite, but
large coupling the renormalization of the density of states is still large,
see Table I. We calculated it as explained in section IID, Eq.~(\ref{N}). At
large coupling the quantity diverges very fast. The limiting value of $%
\widehat{q}$ at zero frequency for various "bare" diffusion constants and
couplings are given in Table II.

\begin{table}[tbp]
\caption{Renormalization of the inverse relaxation time $\widehat{q}$ for
various couplings $r_s$ and bare diffusion constants $\protect\lambda$.}%
\vspace{.2cm}
\begin{tabular}{||c|cccccc||}
\hline\hline
{\large {$_\lambda$}} \hspace{-0.5cm} {\Huge {$\diagdown$}} \hspace{-.5cm}%
{\Large {$^{^{r_{s}}}$}} & $0.1$ & $1$ & $2$ & $4$ & $8$ & $16$ \\ \hline
$8$ & $0.028$ & $0.013$ & $0.0070$ & $0.0034 $ & $0.0015$ & $0.00067$ \\
$4$ & $0.057$ & $0.028$ & $0.015$ & $0.0075$ & $0.0033$ & $0.0015$ \\
$2$ & $0.12$ & $0.056$ & $0.034$ & $0.017$ & $0.0073$ & $0.0032$ \\
$0.5$ & $0.47$ & $0.29$ & $0.18$ & $0.086$ & $0.037$ & $0.015$ \\
$0.1$ & $2.4$ & $1.6$ & $1.3$ & $0.67$ & $0.27$ & $0.10$ \\ \hline\hline
\end{tabular}%
\end{table}

Before trying to further exploit the solutions of the saddle point
equations, we would like to explicitly show that even perturbatively the
reduction in the DOS due to exchange can be clearly seen.

\subsection{Reduction in density of states}

The fact that at small $\varepsilon $ (deviation of fermion's energy from $%
\mu $) and large coupling $r_{s}$, $\Sigma _{\varepsilon }\gg 2\mu
\varepsilon $ leads to significant rearrangement of the DOS in the vicinity
of Fermi surface. The DOS $N(\omega )$ is related to the retarded Green
function by:%
\begin{equation}
N(\omega )=-\frac{1}{\pi }\mathrm{Im}\int_{p}G_{p}^{ret}(\omega ).
\label{Ngen}
\end{equation}%
For frequency independent $q_{\omega }$ it reduces to
\begin{eqnarray}
N(\omega ) &=&-\frac{1}{\pi }\mathrm{Im}\sum_{p}\frac{1}{\omega -2\mu
\varepsilon _{p}-\Sigma _{p}+iq}  \nonumber \\
&=&\frac{1}{2\pi ^{2}}\int_{\varepsilon }\frac{\widehat{q}}{\left( 
\widehat{%
\omega }-\varepsilon -e_{\varepsilon }\right) ^{2}+\widehat{q}^{2}}.
\label{N}
\end{eqnarray}%
\begin{figure}[t]
\centerline{ \psfig{file=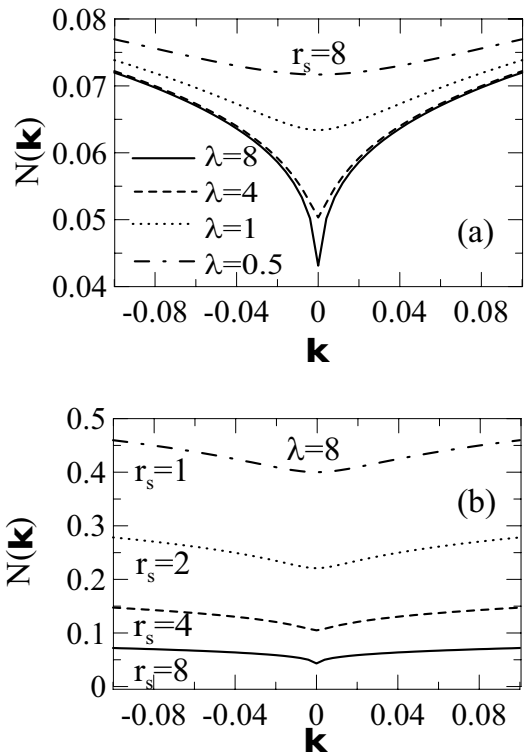,width=0.45\textwidth}}
\caption{The DOS $N(\protect\omega )$ (normalized to the ideal gas DOS $%
N_{0}=1/2\protect\pi $) for (a) fixed $r_{s}=8$ and various 
$\protect\lambda$%
, (b) fixed $\protect\lambda=8$ and various coupling $r_{s}$.}
\label{fig6}
\end{figure}
The DOS at $\omega =0$ for various couplings are given on Table~I, while the
DOS as a function of $\omega $ for fixed $r_{s}=8$ and various $\lambda $ on
Fig. 6a and for fixed $\lambda =8$ and various coupling $r_{s}$ on Fig.~6b.
At large coupling the renormalization of limiting value of $\tau _{r}$ is
very large and $\widehat{q}$ approaches zero. The integrand in Eq.~(\ref{N})
becomes delta function and the DOS is determined by the derivative of $%
e_{\varepsilon }$ with respect to $\varepsilon $ at $\varepsilon =0$:%
\begin{equation}
N(\omega )=\frac{1}{2\pi }\frac{1}{1+e_{\varepsilon }^{\prime }}\Big|_{%
\widehat{\omega }=\varepsilon +\widehat{e}_{\varepsilon }}.  \label{Aapp}
\end{equation}%
Using the simplified saddle point equation Eq.~(\ref{simpE}) the derivative
of $e_{\varepsilon }$ is:%
\begin{equation}
e^{\prime }\equiv \frac{d}{d\varepsilon }e\Big|_{\varepsilon =0}=-\frac{2%
\sqrt{2}r_{s}}{\pi ^{2}}\int_{\varepsilon \geq 0}\kappa ^{\prime
}[\varepsilon ]\left( \frac{\pi }{2}-\arctan \frac{\widehat{q}}{%
e_{\varepsilon }}\right) .  \label{eprimeapp}
\end{equation}%
The asymptotic of the function $\kappa ^{\prime }[\varepsilon ]$ is as
follows. At small $\varepsilon $ it is negative and divergent, $\kappa
^{\prime }[\varepsilon ]\approx -1/2\varepsilon $. As we have seen in the
clean case this lead to vanishing DOS. However in the presence of disorder
the divergence is cut off since the second multiplier vanishes at small $%
\varepsilon $ as $\widehat{e}_{\varepsilon }/\widehat{q}$. This in
particular means that in the case of disorder the RPA improvement is not
necessary as long as the cutoff due to disorder is larger than the cutoff
due to screening (see discussion in subsection IIB).

At large $\varepsilon $ the integral converges rapidly since $\kappa
^{\prime }[\varepsilon ]\approx -\pi /2\varepsilon ^{2}$. This can be seen
perturbatively as well using results of the previous subsection. To leading
order in $r_{s}$%
\[
e^{\prime }=-\frac{2\sqrt{2}r_{s}^{2}}{\pi ^{2}\widehat{q}^{(0)}}%
\int_{\varepsilon \geq 0}\kappa ^{\prime }[\varepsilon ]e_{\varepsilon
}^{(1)},
\]%
where $\widehat{q}^{(0)}$ and $e_{\varepsilon }^{(1)}$ are given in Appendix
Eqs.~(\ref{q0})-(\ref{e1}). The frequency dependent correction to the DOS
(compared to ideal gas $N_{0}=1/2\pi $) is perturbatively
\[
\delta N(\omega )=N(\omega )-N_{0}=\frac{r_{s}}{8\pi ^{2}\lambda ^{2}}%
\int_{\varepsilon }\frac{(\widehat{\omega }-\varepsilon )\widehat{e}%
_{\varepsilon }^{(1)}}{\left[ \left( \widehat{\omega }-\varepsilon \right)
^{2}+\widehat{q}^{(0)2}\right] ^{2}}.
\]%
In particular on the Fermi surface
\[
\delta N(0)=\frac{32r_{s}\lambda ^{3}}{\pi }\widehat{q}^{(1)},
\]%
where $\widehat{q}^{(1)}$ is given in Appendix Eq.~(\ref{q1}). The results
for the DOS are given in Table I. The effect of reduction of the DOS at the
Fermi level due to repulsion increases fast with increasing $r_{s}$ and
depends weaker on $\lambda $. As we mentioned, in the clean case the DOS
approaches zero at any coupling no matter how small, which is clearly an
artifact of neglecting screening at this stage. However the disorder
effectively "averages" the distribution of states making it finite. At the
limit of large disorder the DOS should approach the ideal gas one. Repulsive
interaction works to reduce the DOS near the Fermi surface and is expected
to make scattering by impurities less effective. We will see in the next
section that $e^{\prime }$ is related to the renormalized value of the
diffusion constant: $D_{r}=\lambda /e^{\prime }$.

\section{A systematic expansion in fluctuations around the saddle point.}

\subsection{Classification of fluctuations around the ground state.}

All the Hubbard - Stratonovich fields correspond to elementary (harmonic)
excitations of the system. The spectrum of these excitation is determined by
a quadratic term in expansion of the effective action Eq.~(\ref{Aeff})
around the saddle point determined by a solution of the minimization
equation Eqs.~(\ref{qeq})-(\ref{eeq}). This quadratic form should be
diagonalized to find the spectrum. We first sort out evidently massive modes
which do not play a role in subsequent discussion. These are exchange $E$
and the Cooper exchange $\Theta $ fields. In the next two subsections we
address photons, diffusons and Cooperons.

Since the ground state does not break spontaneously the electric charge $%
U(1) $ symmetry the charged fields $\Theta $ and Cooperon $\Delta $ do not
mix with neutral fields $E$, diffuson $Q$ and static photon $\Phi $. In this
sector the inverse propagator elements (second functional derivatives of the
effective action with respect to relevant fields) is:
\begin{eqnarray}
\left\langle \null_{P^{\prime }p^{\prime }}^{n^{\prime }} \Big| %
D_{\Theta\Theta}^{-1} \Big|\null_{Pp}^{n}\right\rangle &=& \delta
(P-P^{\prime})\delta^{nn^{\prime}} \Big[\delta (p-p^{\prime })v(p)^{-1}
\nonumber \\
&& - \delta (p-p^{\prime })G_{^{P/2+p/2}}^{n}G_{^{P/2-p/2}}^{n}  \nonumber 
\\
&&+\delta(p+p^{\prime })G_{^{P/2+p/2}}^{n}G_{^{P/2-p/2}}^{n}\Big] ,
\nonumber \\
\left\langle \null_{P^{\prime }p^{\prime }}^{n^{\prime }m^{\prime }}\Big| %
D_{\Delta \Theta }^{-1} \Big|\null_{Pp}^{n}\right\rangle &=& \delta
(P-P^{\prime })\delta ^{n^{\prime }+m^{\prime },n}G_{P/2+p/2}^{n^{\prime
}}G_{P/2-p/2}^{m^{\prime }} ,  \nonumber \\
\left\langle \null_{p^{\prime }}^{n^{\prime }m^{\prime }}\Big|D_{\Delta
\Delta ^{\ast }}^{-1}\Big|\null_{p}^{nm}\right\rangle &=& \delta
^{n+m,n^{\prime }+m^{\prime }}\delta (p-p^{\prime })  \nonumber \\
&& \Big(\delta ^{nn^{\prime }}-\frac{1}{\tau }\sum%
\limits_{q}G_{p+q}^{n}G_{q}^{m}\Big) ,  \label{Cooperon}
\end{eqnarray}
where $P$, $p$ are a total and relative momenta of the two electron state
respectively. The relation to indices of the $\Theta $ fields used in
effective action Eq.~(\ref{Aeff}) is obvious%
\[
P=p_{1}+p_{2};\text{ \ \ }p=p_{1}-p_{2}.
\]%
The $\Theta \Theta $ element of the inverse fluctuations propagator
indicated that in this channel there are no massless modes. This is evident
at small coupling since the diagonal first term dominates, but since the
Coulomb interaction is repulsive is true for any coupling. The mixing with
Cooperon cannot turn it to a massless mode. We therefore discard the field 
$%
\Theta $ in what follows. Now we turn to the neutral fields.

Second derivative of the effective action with respect to $E$ is
\begin{eqnarray}
\left\langle\null_{P^{\prime }p^{\prime }}^{n^{\prime 
}}\Big|D_{EE}^{-1}\Big|%
\null_{Pp}^{n}\right\rangle &=& \delta (P-P^{\prime })\delta^{nn^{\prime }} 
\Big( \delta (p-p^{\prime}) v(p)^{-1}  \nonumber \\
& - & \delta (p-p^{\prime}) G_{P/2+p/2}^{n} G_{P/2-p/2}^{n} \Big) .
\end{eqnarray}
It is again a massive mode and its mixing with other neutral field is
unimportant. More interesting are the diffuson $Q$ and the static photon $%
\Phi $ fields:
\begin{eqnarray}
\left\langle \null_{p^{\prime }}^{n^{\prime }m^{\prime }}\Big|D_{QQ}^{-1}%
\Big|\null_{p}^{nm}\right\rangle &=& \delta ^{n-m,n^{\prime }-m^{\prime
}}\delta (p-p^{\prime })  \nonumber \\
&& \left[ \delta ^{nn^{\prime }}-\frac{1}{\tau }\sum%
\limits_{q}G_{p+q}^{n}G_{q}^{m}\right] ,  \label{DmQQ}
\end{eqnarray}%
\begin{equation}
\left\langle \null_{p^{\prime }}^{l}\Big|D_{\Phi Q}^{-1}\Big|\null%
_{p}^{nm}\right\rangle =-\delta ^{l,n-m}\delta (p-p^{\prime })\frac{1}{\tau 
}%
\sum\limits_{q}G_{p+q}^{n}G_{q}^{m} ,  \label{DmQFi}
\end{equation}%
\begin{equation}
\left\langle \null_{p^{\prime }}^{l^{\prime }}\Big|D_{\Phi \Phi }^{-1}\Big|%
\null_{p}^{l}\right\rangle =\delta ^{ll^{\prime }}\delta (p-p^{\prime }) %
\left[ v(p)^{-1}\text{ }-\frac{T}{\tau }\sum_{qn}G_{p+q}^{n}G_{q}^{m}\right]
.  \label{DmFiFi}
\end{equation}
To conclude there are three fields which might have potentially massless
modes. While Cooperon cannot mix with the other two, diffusons and photons
can. Even if both the diffuson and the photon are massless, after mixing the
massless modes could turn massive. We study this phenomenon next.

\subsection{Anderson - Higgs mechanism for diffusons}

\subsubsection{Matrix elements of the photon - diffuson inverse propagator
matrix}

At small frequency and momenta around Fermi surface we will use the
asymptotic expressions for the solution of the saddle point equations%
\begin{eqnarray}
q_{\omega } &\approx &q_{\omega =0}=2\mu \sqrt{\tau } \widehat{q}=\frac{%
\sqrt{\tau }}{2\tau _{r}} ,  \label{renorm} \\
\Sigma _{\varepsilon } &\approx &2\mu e^{\prime }\varepsilon \equiv 2\mu
(Z^{-1}-1)\varepsilon ,  \nonumber
\end{eqnarray}
where renormalizations of the \textquotedblright disorder
efficiency\textquotedblright\ $1/\tau $ and of the inverse density of states
at Fermi level $Z^{-1}$ were introduced in Eq.~(\ref{Z}). As can be seen
from Table~I and Table~II at large coupling they are quite large.

Then one computes standard diagrams in Eq.~(\ref{DmQQ}) for matrix elements
of the inverse propagators involving diffusons and static photons at certain
fixed momentum $p$:%
\begin{eqnarray}
\left\langle n^{\prime }m^{\prime }\Big|D_{QQ}^{-1}\Big|nm\right\rangle
&=&\delta ^{n-m,n^{\prime }-m^{\prime }}\left[ \delta ^{nn^{\prime 
}}-\frac{1%
}{\tau }\sum\limits_{q}G_{p+q}^{n}G_{q}^{m}\right]  \nonumber \\
&\equiv &\delta ^{n-m,n^{\prime }-m^{\prime }}[\delta ^{nn^{\prime }}
\nonumber \\
&&-\theta (-nm)B(p,n-m)],  \label{DmQQ1}
\end{eqnarray}%
where at small $\omega $ and $\ p$, it contains well studied function
\textquotedblright bubble\textquotedblright\ integral having the following
asymptotic at small frequencies and momenta:
\begin{equation}
B(p,l)=\frac{1}{\tau }\sum\limits_{q}G_{p+q}^{n}G_{q}^{n+l}\simeq 1-\omega
_{l}\tau _{r}-D_{r}p^{2}.  \label{bubble}
\end{equation}%
for $n<0,$ $n+l>0.$ Within an approximation of \textquotedblright
renormalization\textquotedblright\ Eq.~(\ref{renorm}) one obtains:
\[
D_{r}=\frac{Z\tau _{r}^{3}}{\tau ^{3}}\lambda .
\]%
Beyond this approximation the value of effective diffusion constant can be
calculated from the values given in Tables I and II. As usual, we discard
the massive same frequency sign $++$ and $--$ modes \cite{Belitz} and
concentrate on different frequency sign excitations.

It is convenient to this end to rescale the photon field%
\begin{equation}
\Phi _{pn}=\sqrt{v(p)T}\widetilde{\Phi }_{pn}.  \label{phirescale}
\end{equation}%
The mixing and the photon inverse propagators are:
\begin{equation}
\left\langle \null_{p^{\prime }}^{l}\Big|D_{\Phi Q}^{-1}\Big|\null%
_{p}^{nm}\right\rangle _{l}=-\delta ^{l,n-m}\delta (p-p^{\prime 
})\sqrt{v(p)T%
}\theta (-nm)B(p,n-m),  \label{DmFiQ1}
\end{equation}%
\begin{equation}
\left\langle \null_{p^{\prime }}^{l^{\prime }}\Big|D_{\Phi \Phi }^{-1}\Big|%
\null_{p}^{l}\right\rangle _{^{l^{\prime }}}=\delta ^{ll^{\prime }}\delta
(p-p^{\prime })\left[ 1\text{ }-v(p)L(p,l)\right] ,  \label{DmFiFi1}
\end{equation}%
where the Lindhard function \cite{Giuliani} has an asymptotic behaviors:
\begin{equation}
L(p,l)\equiv T\sum_{n}B(p,n,n-l)\approx (-1+\omega _{l}\tau _{r}+...)/(2\pi
).  \label{J}
\end{equation}%
One notices in Eq.~(\ref{DmFiQ1}) that, due to the factor $\theta (-nm),$ it
is precisely the massless different frequency sign diffusons that mix with
photon. The mixing strength is large since it is determined by the same
bubble integral that appears in the diffuson's inverse propagator Eq.~(\ref%
{DmQQ1}). Now we will invert this matrix and find its eigenmodes.

\subsubsection{Eigenvalues and eigenmodes: physical photon and diffuson}

The $Q\Phi $ inverse propagator matrix is \textquotedblright
blocked\textquotedblright\ for different photon frequencies $l=n-m$ (we take
$n>0$ $m<0$), namely we can consider a single value of $l$. For a fixed
frequency $l>0$ the range of possible $n$ is limited to $-l<n<0$ and the $%
(l+1)\times (l+1)$ matrix has the following form:
\begin{equation}
\left(
\begin{array}{cccc}
a & 0 & \ldots & b \\
0 & a & \ldots & b \\
\vdots & \vdots & \ddots & \vdots \\
b & b & \ldots & c%
\end{array}%
\right) ,  \label{invmatrix}
\end{equation}%
where
\begin{equation}
a=1-B(p,l),b=-\sqrt{v(p)T}B(p,l),c=1-v(p)TL(p,l).  \label{abc}
\end{equation}%
Eigenvalues of this matrix are: the $(l-1)$ times degenerate $a\sim \omega
_{l}\tau _{r}+$ $D_{r}p^{2}$ (original \textquotedblright
massless\textquotedblright\ diffusons before mixing) and two nondegenerate
eigenvalues
\[
\lambda _{\pm }=\frac{1}{2}\left( a+c\pm \sqrt{(a-c)^{2}+4lb^{2}}\right) ,
\]%
corresponding to eigenvectors $\{1,1,1...1,a_{\pm }\}$ with $a_{\pm
}=(\lambda _{\pm }-a)/b.$ Their asymptotic at small momenta and frequency
is:
\[
\lambda _{+}\approx 1+v(p)/(2\pi ),
\]%
\begin{equation}
\lambda _{-}\approx 2\pi v(p)^{-1}(\omega _{l}\tau
_{r}+D_{r}p^{2})+D_{r}p^{2}.  \label{lambda}
\end{equation}%
The form of the $\lambda _{+}$ eigenvalue means that physical photon mode $%
\overline{\Phi }_{p}^{\omega }=\sum_{n=-l}^{0}$ $Q_{p}^{n}+a_{+}\Phi
_{p}^{\omega }$ propagator (rescaling back by $\sqrt{v(p)T}$) is the RPA
photon propagator exhibiting Debye screening:
\begin{equation}
\left\langle \overline{\Phi }_{p}^{\omega }\overline{\Phi }_{-p}^{-\omega
}\right\rangle \sim \frac{1}{v^{-1}(p)+2\pi }.  \label{photon}
\end{equation}%
The second is a "symmetric" in $n$ superposition of diffuson modes $%
\overline{Q}_{p}^{\omega }=\sum_{n=-l}^{0}Q_{p}^{n}+a_{-}\Phi _{p}^{\omega 
}$
is no longer the usual diffusion pole:
\begin{eqnarray}
\left\langle \overline{Q}_{p}^{\omega }\overline{Q}_{-p}^{-\omega
}\right\rangle &\sim &\frac{1}{\omega \tau
_{r}+D_{r}p^{2}+D_{r}v(p)p^{2}/(2\pi )}  \label{diffuson} \\
&\simeq &\frac{1}{\omega \tau _{r}+D_{r}e^{2}p/(4\pi )}.
\end{eqnarray}%
It becomes harder (less singular) than in the standard treatment in which
the mixing between photon and diffuson is neglected. It is interesting to
note that in the standard treatment at small couplings the mixing is not
neglected as far as photon's propagator is concerned. One uses the RPA
propagator Eq.~(\ref{photon}), despite the fact that the same off diagonal
matrix element Eq.~(\ref{DmQFi}) is simultaneously responsible for the
essential modification of the diffuson.

The inverse of a matrix of the type of Eq.~(\ref{invmatrix}) generally is:
\[
\frac{1}{ac-lb^{2}}\left(
\begin{array}{cccc}
c-\frac{lb^{2}}{a}+\frac{b^{2}}{a} & \frac{b^{2}}{a} & \ldots & -b \\
\frac{b^{2}}{a} & c-\frac{lb^{2}}{a}+\frac{b^{2}}{a} & \ldots & -b \\
\vdots & \vdots & \ddots & \vdots \\
-b & -b & \ldots & a%
\end{array}%
\right) .
\]%
Matrix elements of the propagators therefore are:
\begin{equation}
\left\langle n,n+l|D_{QQ}|n^{\prime },n^{\prime }+l\right\rangle
=P_{1}\delta _{nn^{\prime }}+P_{2}\frac{1}{2\pi l},  \label{DQQ}
\end{equation}%
\begin{equation}
\left\langle n,n+l|D_{Q\Phi }|l\right\rangle =P_{3}\frac{1}{\sqrt{2\pi l}},
\label{DQFi}
\end{equation}%
\begin{equation}
\left\langle l|D_{\Phi \Phi }|l\right\rangle =P_{4},  \label{DFiFi}
\end{equation}%
where functions $P$ and their asymptotic at small frequency and momentum
are:
\begin{eqnarray}
P_{1} &=&\frac{1}{a}\approx \frac{1}{\tau _{r}\omega +D_{r}p^{2}},
\label{pp1} \\
P_{2} &=&\frac{lb^{2}}{a(ac-lb^{2})}\approx \frac{v(p)}{\tau _{r}\omega
+D_{r}p^{2}+D_{r}p^{2}v(p)}\bigskip ,  \label{pp2} \\
P_{3} &=&-\frac{\sqrt{2\pi l}b}{ac-lb^{2}}\approx \frac{\sqrt{\omega 
v(p)}}{%
\tau _{r}\omega +D_{r}p^{2}+D_{r}p^{2}v(p)},  \label{pp3} \\
P_{4} &=&\frac{a}{ac-lb^{2}}\approx \frac{\tau _{r}\omega +D_{r}p^{2}}{\tau
_{r}\omega +D_{r}p^{2}+D_{r}p^{2}v(p)}.  \label{pp4}
\end{eqnarray}

We will see in subsection~IIID that the diagonal part of the diffuson
describes density fluctuations and therefore mixing with photon makes
electrons nondiffusive at large densities.

\begin{figure}[t]
\centerline{\psfig{file=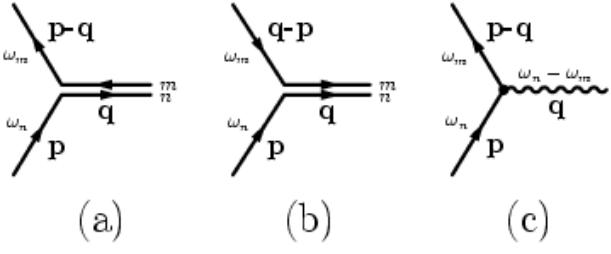,width=0.5\textwidth}}
\caption{The Feynman rules for (a) fermion-fermion-diffuson, (b)
fermion-fermion-Cooperon, (c) fermion-fermion-photon vertex. The solid and
wavy lines denote the fermion and photon propagator respectively, the double
solid lines with opposite (same) direction arrows denote the diffuson
(Cooperon) propagator.}
\label{fig7}
\end{figure}

Propagators for relevant modes supplemented by vertices, the fermion -
fermion - diffuson
\[
\Gamma _{\overline{\psi }\psi Q}=-\frac{i}{\tau }
\]%
and the fermion - fermion - photon $\Phi $%
\[
\Gamma _{\overline{\psi }\psi \Phi }=-i
\]%
constitute Feynman rules shown on Fig. 7(a) and (c) respectively. However
fields $\Phi $ and $Q$ do not correspond to \textquotedblright
modes\textquotedblright\ or \textquotedblright bosonic
excitations\textquotedblright\ of the system due to mixing between them
discussed in detail in the previous subsection. One can still use these
fields in calculation considering them as a vector and their propagator as a
matrix.

\subsection{The Cooperon propagator}

\begin{figure}[b]
\centerline{ \psfig{file=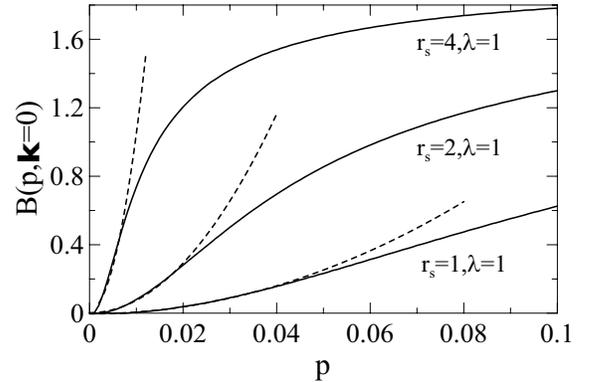,width=0.45\textwidth}}
\caption{The dependence of $B(p,\omega=0)$ on $p$ for various $r_s$
and $\lambda$. The solid lines are the original values of
 $B(p,\omega=0)$%
, while the dashed lines are their corresponding quadratic fitting values.}
\label{fig9}
\end{figure}

As we mentioned in section IIIA due to its charge the Cooperon does not mix
with photon or any other neutral field. It is massless for different sign
frequencies and massive for the same sign frequencies. The strong coupling
however influences its propagator beyond the evident renormalization $%
\lambda \rightarrow D_{r}$, $\tau \rightarrow \tau _{r}$. Substituting the
expression of the fermion propagators Eq.~(\ref{Gm1}) into Eq.~(\ref%
{Cooperon}) the inverse propagator of the excitation is:%
\begin{equation}
\left\langle nm|D_{\Delta \Delta ^{\ast }}^{-1}|nm\right\rangle =1-\theta
(-nm)B(p,n-m).  \label{DmDelDel1}
\end{equation}%
Numerical solution of the saddle point equations substituted into Eq.~(\ref%
{Cooperon}) show that the dependence is quadratic only till certain momentum
at which it saturates, see Fig.~\ref{fig9} for $r_{s}=1,2$ and $4$. This is
of importance later when we estimate the quantum correction to conductivity
in section IV.

Note that the excitation remains massless even at strong coupling. This
follows from very general considerations. Consider the \textquotedblright
bubble\textquotedblright\ diagram
\begin{equation}
B(\omega _{l},p)=\frac{1}{\left( 2\pi \right) ^{2}}\int_{q}G_{qn}G_{q+p,m},
\label{bubble1}
\end{equation}%
where $l=m-n$. At zero momentum using the saddle point equation Eq.~(\ref%
{SPq}) one obtains:
\begin{eqnarray}
B(\omega _{l},p=0) &=&\frac{1}{\left( 2\pi \right) ^{2}}%
\int_{q}(G_{qn}-G_{qm})\frac{1}{i\left[ \omega _{l}+q_{m}-q_{n}\right] }
\nonumber \\
&=&\frac{q_{m}-q_{n}}{\omega _{l}+q_{m}-q_{n}}.  \label{bub0}
\end{eqnarray}%
This approaches $1$ in the limit of zero frequency as long as there is a
jump in $Q_{\omega }$ from negative to positive frequencies. The propagator
of the different frequency sign Cooperons at small frequencies and momenta
is:%
\begin{equation}
D_{\Delta \Delta ^{\ast }}=\frac{1}{2q\omega +B^{\prime }p^{2}},
\label{DelDel}
\end{equation}%
where $B^{\prime }$ denotes a derivative of the bubble integral
\begin{equation}
B^{\prime }=\frac{\partial B(0,p)}{\partial p^{2}}\Big|_{p=0}.
\label{bderiv}
\end{equation}%
Assuming quadratic momentum dependence of $B(p,\omega )$, direct calculation
leads to
\begin{equation}
B^{\prime }=\frac{1}{2\pi (2\mu)^2 }\int\limits_{\varepsilon }\frac{%
(\varepsilon +e_{\varepsilon }^{\prime })^{2}-\widehat{q}^{2}}{\left[
\widehat{q}^{2}+(\varepsilon +e_{\varepsilon }^{\prime })^{2}\right] ^{3}}%
(1+e_{\varepsilon }^{\prime })^{2}.  \label{bubp}
\end{equation}

\begin{widetext}

\begin{table}[t]
\caption{Derivative $B^{\prime }/tau^2$ of the Lindhard function in the
presence of disorder for various values of $r_{s}$ and
$\lambda$}
\vspace{.2cm}
\begin{tabular}{||c|cccccc||}
\hline\hline {\large {$_\lambda$}} \hspace{-0.5cm} {\Huge
{$\diagdown$}} \hspace{-.5cm}{\Large {$^{^{r_{s}}}$}} & $0.1$ &
$1$ &$2$ & $4$ & $8$ & $16$  \\ \hline
$8$ & $12.2$ & $3.13\times 10^2$ & $3.33\times 10^3$ &$6.06\times 10^4$ & 
$1.40\times 10^{6}$ & $3.12\times 10^{7}$ \\
$4$ & $5.78$ & $1.14\times 10^2$ &$1.15\times 10^3$&$2.11\times 10^4$& 
$5.14\times 10^5$ & $1.29\times 10^{7}$ \\
$2$ & $2.74$ & $40.4$&$3.77\times 10^2$ &$7.02\times 10^3$ & $1.81\times 
10^5$ & $4.93\times 10^{6}$  \\
$0.5$ & $0.620$ & $4.75$ &$34.8$ &$6.35\times 10^2$ & $1.90\times 10^4$ & 
$6.18\times 10^5$\\
$0.1$ & $0.117$ & $0.605$ &$1.79$& $25.3$ & $9.15\times 10^2$ & $4.10\times 
10^4$  \\
\hline\hline
\end{tabular}%
\end{table}

\end{widetext}

The values of $B^{\prime }$ for various couplings $r_{s}$ and bare diffusion
constants $\lambda $ determining the Drude conductivity are given in Table
III. The perturbative expression for this quantity is
\begin{equation}
B^{\prime }=-\lambda \tau^2 \left[ 1+r_{s}\Big(-12\lambda \widehat{q}^{(1)}+\frac{%
2^{5/2}\lambda ^{3}}{\pi ^{2}}C\Big)\right] ,  \label{Bppert}
\end{equation}%
where
\begin{eqnarray}
C &=&\frac{1}{(4\lambda )^{4}}G_{33}^{22}\left( (4\lambda )^{2}\Big|\null%
_{1,1}^{3/2,3/2}\right) +G_{43}^{23}\left( \frac{1}{(4\lambda )^{2}}\Big|%
\null_{3/2,3/2,3}^{2,2,2}\right)  \nonumber \\
&&+2G_{43}^{23}\left( \frac{1}{(4\lambda )^{2}}\Big|\null%
_{3/2,3/2,4}^{2,2,2}\right) ,  \label{C}
\end{eqnarray}%
with the standard notation $G_{pq}^{mn}\left( z\Big|\null%
_{b_{1},...,b_{q}}^{a_{1},...,a_{p}}\right) $ of the Meijer functions. 
\cite%
{Gradstein} It rises fast with $r_{s}$ , while being weakly dependent on $%
\lambda $. Finally the fermion - fermion - Cooperon vertex is
\[
\Gamma _{\overline{\psi }\psi \Delta }=-\frac{i}{\tau }
\]%
and it completes the Feynman rules shown on Fig.~7(b).

\subsection{Density - density correlator and conductivity to leading order}

\subsubsection{Modification of diffusive motion due to strong interaction}

\begin{figure}[t]
\centerline{ \psfig{file=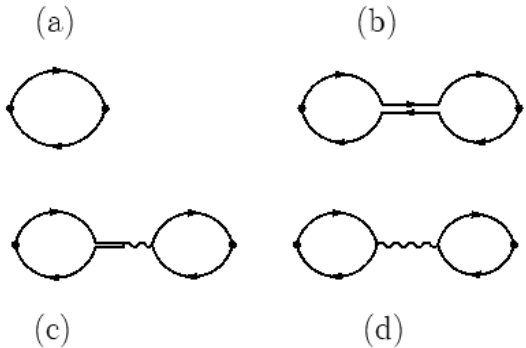,width=0.4\textwidth}}
\caption{Contributions to the density-density correlator at the leading
order.}
\label{fig10}
\end{figure}

One of the most important characteristics of 2DEG is the density - density
correlator describing the diffusive nature of the charge carrier's motion in
a disordered medium. It is closely related to dielectric function and
polarizability of 2DEG. The correlator is given in Matsubara formalism by
\begin{eqnarray}
&\chi (\omega ,p) = &  \nonumber \\
&\sum\limits_{q_{1},q_{2}}\int\limits_{0}^{1/T} d\tau e^{i\omega \tau } &
\left\langle T_{\tau }\left[ \overline{\psi }_{p+q_{1}}(\tau )\psi
_{q_{1}}(\tau )\overline{\psi }_{p+q_{2}}(0)\psi _{q_{2}}(0)\right]
\right\rangle .  \nonumber
\end{eqnarray}
First we use the Feynman rules stated above to calculate the density -
density correlator at the leading order. In the limit of small frequencies
the contributions come from diagrams (a)-(d) on Fig.~\ref{fig10}. They are 
$%
-L(p,l)$, $-\omega _{l}B(p,l)^{2}P_{1}(1+P_{4})/2\pi $, $2\sqrt{\omega
_{l}v(p)/2\pi }L(p,l)B(p,l)P_{3}$, and $-v(p)L(p,l)^{2}P_{4}$, respectively
and can be combined into
\[
\chi (\omega _{l},p)=\frac{\chi _{0}(\omega _{l},p)}{1+v(p)\chi _{0}(\omega
_{l},p)}.
\]%
Here expressions for $L,B$ and $P_{4}$ are given in Eqs.~(\ref{J}), (\ref%
{bubble1}), (\ref{pp1})-(\ref{pp4}) and the \textquotedblright
noninteracting\textquotedblright\ correlator is defined by
\[
\chi _{0}(\omega _{l},p)=-\left( L(p,l)+\frac{\omega _{l}}{2\pi }\frac{%
B(p,l){}^{2}}{1-B(p,l)}\right) .
\]%
Its asymptotic at small $\omega $ and $p$ is:%
\[
\chi (\omega _{l},p)=\frac{D_{r}p^{2}}{\omega_l 
\tau_{r}+D_{r}p^{2}+D_{r}p^{2}v(p)/(2\pi )}.
\]%
Therefore not surprisingly it is proportional to propagator of the
\textquotedblright diagonal\textquotedblright\ diffuson defined in 
Eq.~(\ref{diffuson}). The diffusive behavior dominates short range fluctuations only
on scale smaller than $s=2\epsilon /e^{2}.$ On a larger scale the last term
in the denominator is linear in $p$ and is therefore larger than the
standard diffusion term. This makes diffusion less long range although in 2D
it does not become a short range one. The scale was introduced by Si and
Varma \cite{Si} and we will comment on connection to their work in section 
V.

\subsubsection{The leading (Drude) contribution conductivity}

\begin{table}[b]
\caption{The Drude conductivity $\protect\sigma/2 \protect\pi e^2$ for
various values of $r_s$ and $\protect\lambda$.}\vspace{.2cm}
\begin{tabular}{||c|cccccc||}
\hline\hline
{\large {$_\lambda$}} \hspace{-0.55cm} {\Huge {$\diagdown$}} \hspace{-.5cm}%
{\Large {$^{^{r_{s}}}$}} & $0.1$ & $1$ & $2$ & $4$ & $8$ & $16$ \\ \hline
$8$ & $9.85$ & $51.4$ & $1.68\times 10^2$ & $7.15\times 10^2$ & $3.39\times
10^3$ & $1.53\times 10^4$ \\
$4$ & $4.79$ & $22.0$ & $70.3$ & $3.01\times 10^2$ & $1.47\times 10^3$ & $%
7.18\times 10^3$ \\
$2$ & $2.33$ & $9.31$ & $28.7$ & $1.24\times 10^2$ & $6.22\times 10^2$ & $%
3.20\times 10^3$ \\
$0.5$ & $0.553$ & $1.61$ & $4.44$ & $19.0$ & $1.03\times 10^2$ & $5.77\times
10^2$ \\
$0.1$ & $0.107$ & $0.246$ & $0.470$ & $1.79$ & $10.5$ & $68.3$ \\
\hline\hline
\end{tabular}%
\end{table}

The DC conductivity can be read off the density - density correlator using
the relation
\begin{eqnarray}
\sigma &=&\lim_{\omega \rightarrow 0}\lim_{p\rightarrow 0}\frac{e^{2}\omega
}{p^{2}}\chi (\omega ,p)  \nonumber \\
&=&-\lim_{\omega \rightarrow 0}\frac{e^{2}\omega ^{2}}{2\pi }\frac{B(\omega
,0)^{2}}{\left( 1-B(\omega ,0)\right) ^{2}}B^{\prime }  \nonumber \\
&=&\frac{e^{2}}{2\pi }4q^{2}B^{\prime }/\tau,  \label{drude}
\end{eqnarray}%
which follows from the Kubo formula. \cite{Vollhardt} Here we used the
asymptotic of the bubble integral $B$ Eq.~(\ref{J}) and the imaginary part
of self energy $q$ and derivative of the bubble integral $B^{\prime }$
defined in Eqs.~(\ref{SPq}) and (\ref{bderiv}) respectively. Results are
given in Table IV for various $r_{s}$ and $\lambda $. One observes that at
large coupling the \textquotedblright Drude\textquotedblright\ conductivity
increases considerably compared with the noninteracting one. We explain this
by reduction of interaction with disorder ($q$ is much smaller than its
noninteracting value of $1/2\tau $) despite the reduction in density of
states ($B^{\prime }$ larger than its noninteracting value of $-\mu $). At
small $r_{s}$ using \ Eqs.~(\ref{e1})-(\ref{q1}) of Appendix and Eq.~(\ref%
{Bppert}) one obtains%
\[
\sigma =\frac{e^{2}}{2\pi }\lambda \left( 1-4r_{s}\lambda \widehat{q}%
^{(1)}+r_{s}\frac{2^{5/2}\lambda ^{3}}{\pi ^{2}}C\right) ,
\]%
where $C$ is given in Eq.~(\ref{C}). The leading order contribution
dominates at small couplings and disorder. However, the theory has zero
modes - Cooperons. Therefore possible IR divergencies might render the
leading order results invalid at large coupling or disorder. In principle
for zero temperature and infinite samples the results are invalid for all
couplings. Our next task is find a range of parameters and temperature (or
sample sizes) in which the IR divergencies at the next order are still small
compared to the main contribution.

\section{Suppression of weak localization by the long range interaction
effects}

\subsection{The saddle point expansion and the spin-singlet approximation}

In this section we describe some of the corrections around the variational
ground state found in section II and used in section III to calculate
several physical quantities. The steepest descent expansion in terms of
Feynman diagrams is quite standard. \cite{Wegner,Belitz} We briefly describe
it introducing $N_{s}$ identical \textquotedblright spin\textquotedblright\
components to show that the expansion might be interpreted as
\textquotedblright $1/N_{s}$\textquotedblright\ expansion (spin might
include other degeneracies like multiple valleys in Si). The action
including the spin indices $\sigma $ is given in Eq.~(\ref{Aspin}). It is a
peculiar feature of the disordered Coulomb problem that the leading order in
$1/N_{s}$ vanishes for two entirely unrelated reasons. The direct
contribution in the disorder part vanishes due to the fact that it is of
higher (second) order in replicas $N_{r}$ as well:%
\[
\sum_{a,b}\sum_{\sigma _{1,}\sigma _{2}}G_{\sigma _{1}\sigma
_{1}}^{aa}G_{\sigma _{2}\sigma _{2}}^{bb}\sim N_{r}^{2}N_{s}^{2}.
\]%
The direct $N_{s}^{2}$ contribution to the Coulomb part
\[
\int_{x,y}\sum_{a}\sum_{\sigma _{1,}\sigma _{2}}G_{\sigma _{1}\sigma
_{1}}^{aa}(x,x)v(x-y)G_{\sigma _{2}\sigma _{2}}^{aa}(y,y)=0
\]%
vanishes due to neutralizing background $\int_{y}v(x-y)=0$ (and under
assumption of homogeneity)$.$ Therefore leading terms are of order $N_{s}.$
The free theory action is also of order $N_{s}$. Therefore all the terms in
action are of the order $N_{s}$ and it plays a role of the
\textquotedblright loop expansion parameter\textquotedblright\ and comes
always in combination with $1$/$\hbar$. This Hartree - Fock logic is not
rigorous however. There is an assumption involved: it is assumed that all
the Hubbard - Stratonovich fields which in general are tensorial are
dominated by their singlet part:%
\[
Q^{\sigma _{1}\sigma _{2}}\sim \delta ^{\sigma _{1}\sigma _{2}}Q.
\]%
In this paper we will make such an assumption. Therefore we neglect, for
example, triplet channels in the physical case of $N_{s}=2.$

The partition function (suppressing for simplicity the replica indices and
writing explicitly just one of the HS fields) or any observable is expanded
around the saddle point
\begin{eqnarray*}
Z &=&\int_{Q}e^{N_{s}A_{eff}[Q]}  \nonumber \\
& \approx & \int_{Q}\exp [N_{s}A_{eff}[Q_{SP}]+\frac{1}{2}%
QD^{-1}[Q_{SP}]Q+\Delta A[Q]],
\end{eqnarray*}
where $\Delta A[Q]$ contains all the cubic, quartic and higher order terms
in $Q$. From this Feynman rules are read and they scale compared to 
Fig.~\ref%
{fig7} in the following way: HS fields' propagators are proportional to $%
N_{s},$, fermion loop also has $N_{s},$ while fermion - fermion - boson
vertex is $1/\sqrt{N_{s}}$ The leading order contribution of conductivity
considered so far is of order $N_{s}$ and we will consider order $%
N_{s}^{0}=1 $ in the next section.

\subsection{Fluctuation correction to the density - density correlator and
conductivity}

\subsubsection{Density - density correlator}

\begin{figure}[t]
\centerline{ \psfig{file=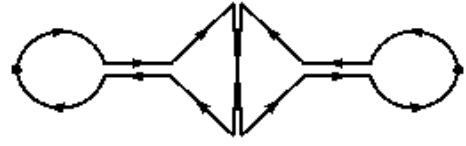,width=0.35\textwidth}}
\caption{The correction to the density-density correlator at two loop 
order.}
\label{fig11}
\end{figure}

The correction to density - density correlator at two loop order which
contributes to the small frequency limit (the only ones needed for
subsequent calculation of the conductivity) is given on Fig.~\ref{fig11}
\begin{eqnarray*}
\delta \chi (\omega _{l},p) &=&T\sum\limits_{q,r}\sum\limits_{nmn^{\prime
}m^{\prime }}\theta (-n(n+l))\theta (-n^{\prime }(n^{\prime }+l))B(p,l)^{2}
\\
&&\left\langle n,n+l|D_{QQ}|m,m+l\right\rangle
G_{q}^{m}G_{p+q}^{m+l}G_{q+r}^{m^{\prime }}G_{p+q+r}^{m^{\prime }+l} \\
&&\left\langle n,m^{\prime }|D_{\Delta \Delta r}|m+l,m^{\prime
}+l\right\rangle \\
&&\left\langle m^{\prime },m^{\prime }+l|D_{QQ}|n^{\prime },n^{\prime
}+l\right\rangle .
\end{eqnarray*}%
All the other diagrams are regular as $\omega \rightarrow 0$ , hence they do
not give contributions to the DC conductivity. Near the Fermi surface one
\textquotedblright disentangles\textquotedblright\ the momenta flowing in
the central loop, see Fig.~\ref{fig11},
\begin{equation}
\delta \chi =T\sum\limits_{n=0}^{l}\frac{B(p,l)^{2}}{(1-B(p,l))^{2}}%
B_{4}(p,l)\sum\limits_{r}\frac{1}{1-B(r,l)},  \label{div}
\end{equation}%
where
\begin{equation}
B_{4}(p,l)\equiv
\sum\limits_{q}G_{q}^{n}G_{-q}^{n+l}G_{p-q}^{n}G_{q-p}^{n+l}.  
\label{bub4}
\end{equation}%
The integral over $r$ is logarithmically infrared divergent in 2D and, as
usual, \cite{Vollhardt} signals breakdown of naive perturbation theory and
appearance of weak localization effects. We assume an IR cutoff (to be
defined more explicitly below) and will use this expression to calculate
conductivity.

\subsubsection{Weak localization}

The fluctuation correction to conductivity using Kubo formula is:
\[
\delta \sigma =\lim_{\omega \rightarrow 0}\frac{e^{2}\omega ^{2}}{2\pi }%
\frac{B(0,l)^{2}}{\left( 1-B(0,l)\right) ^{2}}B_{4}^{\prime}\sum\limits_{r}%
\frac{1}{1-B(r,l)},
\]%
where the derivative of $B_{4}$ Eq.~(\ref{bub4}) is defined by
\[
B_{4}^{\prime}\equiv \frac{\partial B_{4}(p,l=0)}{\partial (p^{2})} \bigg|%
_{p=0}.
\]%
After some algebra it takes a form
\[
B_{4}^{^{\prime }}=\frac{1}{2\pi (2\mu)^4 }\int_{\varepsilon }\frac{(\varepsilon
+e_{\varepsilon })^{2}}{\left( \widehat{q}^{2}+(\varepsilon +e_{\varepsilon
})^{2}\right) ^{4}}(1+e_{\varepsilon }^{\prime })^{2} .
\]%
Values of the coefficient $B_{4}^{\prime}$ for various couplings and
disorder strength are given in Table V.

\begin{widetext}
\begin{center}
\begin{table}[t]
\caption{The values of $B_{4}^{\prime }/\tau^4$ for various $r_s$ and 
$\lambda$}\vspace{%
.2cm}
\begin{tabular}{||c|cccccc||}
\hline\hline {\large {$_\lambda$}} \hspace{-0.5cm} {\Huge
{$\diagdown$}} \hspace{-.5cm}{\Large {$^{^{r_{s}}}$}} & $0.1$ &
$1$ &$2$&$4$& $8$ & $16$ \\ \hline
$8$ & $3.77$ & $4.64\times 10^2$ &$1.60\times 10^4$&$1.25\times 10^6$ & 
$1.47\times 10^{8}$ & $2.11\times 10^{10}$\\
$4$ & $3.46$ & $2.86\times 10^2$ &$9.05\times 10^3$& $7.21\times 10^5$& 
$8.92\times 10^{7}$ & $1.28\times 10^{10}$ \\
$2$ & $3.18$ & $1.70\times 10^2$ &$4.78\times 10^3$ &$3.85\times 10^5$& 
$5.14\times 10^{7}$ & $7.78\times 10^{9}$  \\
$0.5$ & $2.73$ & $53.6$ &$1.04\times 10^3$ &$8.10\times 10^4$& $1.35\times 
10^{7}$ & $2.58\times 10^{9}$ \\
$0.1$ & $2.44$ & $27.4$ &$1.23\times 10^2$ &$6.65\times 10^3$&
$1.51\times 10^{6}$ & $4.72\times 10^{8}$  \\ \hline\hline
\end{tabular}%
\end{table}
\end{center}
\end{widetext}

Before discussing physical implications of the correction we provide
perturbative result for $B_{4}^{\prime }$.
\[
B_{4}^{^{\prime }}=2 \lambda \tau^4 \left[ 1+r_{s}\lambda 
\Big(-2^{2}5\widehat{q}^{(1)}+%
\frac{2^{5/2}\lambda ^{2}}{3\pi ^{2}}C_{4}\Big)\right] ,
\]%
where
\begin{eqnarray*}
C_{4} &=&\frac{3}{2^{7}\lambda ^{4}}G_{33}^{22}\left( (4\lambda )^{2}\Big|%
\null_{1,1}^{3/2,3/2}\right) -9G_{43}^{23}\left( \frac{1}{(4\lambda )^{2}}%
\Big|\null_{3/2,3/2,3}^{2,2,2}\right) \\
&+&12G_{34}^{23}\left( \frac{1}{(4\lambda )^{2}}\Big|\null%
_{3/2,3/2,4}^{2,2,2}\right) +4G_{43}^{23}\left( \frac{1}{(4\lambda )^{2}}%
\Big|\null_{3/2,3/2,5}^{2,2,2}\right) .
\end{eqnarray*}%
The quantity is increasing very fast with coupling $r_{s}$ and decreases
slowly with disorder strength.

Returning to conductivity one obtains%
\[
\delta \sigma =\frac{e^{2}}{2\pi }\frac{2q^{2}B_{4}^{\prime}}{\pi\tau }%
\int_{0}^{\infty }\frac{rdr}{1-B(r,l)}.
\]%
In 2D the integral is dominated by small momenta. Therefore at very low
temperature we can use approximation of Eq.~(\ref{DelDel})%
\begin{equation}
\delta \sigma =\frac{e^{2}}{2\pi }\frac{2q^{2}B_{4}^{\prime}}
{\pi\tau B^{\prime}}
\int_{p_{IR}}^{p_{UV}}\frac{rdr}{r^{2}}=\frac{e^{2}}{2\pi }\frac{%
q^{2}B_{4}^{\prime}}{\pi B^{\prime }}\log \frac{p_{UV}^{2}}{p_{IR}^{2}}.
\label{corr}
\end{equation}%
As usual \cite{Vollhardt} it is cut off in both infrared and ultraviolet.
The infrared cutoff for the weak localization logarithmic divergence can be
set by finite temperature%
\[
p_{IR}^{2}=\frac{2\pi T m^{\ast}/\tau}{2 (q/\sqrt{\tau}) (B^{\prime }/\tau^2)}
\]%
or finite size $L$ of the sample%
\[
p_{IR}=\frac{2\pi }{L}.
\]%
The ultraviolet cutoff is \cite{Sadovsky}
\begin{equation}
\frac{p_{UV}^{2}}{2m^{\ast }}=\min \{\mu ,1/\tau \}.  \label{UV}
\end{equation}

\subsection{Crossover temperature}

Let us find a temperature at which the perturbation theory in
\textquotedblright loops\textquotedblright\ or $1/N_{s}$ breaks down. At
this temperature the Drude conductivity is significantly reduced by
fluctuations and one conservatively estimates it as settling of the weak
localization (the Anderson insulator) regime. It is estimated by equating
leading (Eq.~(\ref{drude})) and the fluctuation correction (Eq.~(\ref{corr}%
)) to conductivity times a factor $R$ of order $1$ at finite temperature (or
sample size):
\begin{eqnarray}
\sigma _{0} &=&\frac{e^{2}}{2\pi }4N_{s}q^{2}B^{\prime }/\tau=R\text{ }\delta
\sigma  \nonumber \\
&=&\frac{e^{2}}{2\pi }\frac{q^{2}B_{4}^{\prime }}{\pi\tau B^{\prime }}\log 
\left[
\left( \frac{2m^{\ast }}{\tau }\right) /\left( \frac{2\pi T_{wl}
m^{\ast}/ \tau}
{2 (q/\sqrt{\tau}) (B^{\prime }/\tau^2)} \right) \right] ,
\end{eqnarray}%
where we used the large disorder value in Eq.~(\ref{UV}). Therefore
\[
T_{wl}=\frac{2 q B^{\prime }}{\pi \tau^{3/2}}e^{-\varsigma },
\]%
with a dominant argument of the exponential being%
\[
\varsigma =\frac{4\pi N_{s}(B^{\prime })^{2}}{RB_{4}^{\prime }}.
\]%
Considering first the bare diffusion constant as fixed one observes that as 
$%
r_{s}$ increases the temperature first rises due to preexponential factor,
but then after reaching a maximum exponentially drops at large $r_{s}$. On
the other hand fixing $r_{s}$ the temperature quickly drops as $\lambda $
increases. In experiment what is usually varied is density of electrons. In
this case as density gets lower both the diffusion constant becomes smaller
and $r_{s}$ becomes larger. The overall effect is that for clean samples and
relatively large $r_{s}$ the quasimetallic state is stable to very long
temperatures due to reduction of the DOS at the Fermi level. Note that the
trajectory in the $\lambda ,r_{s}$ space of the experimental setup is itself
dependent on the DOS. \cite{Davis} This complicates the actual comparison
since effects of screening cannot be neglected in experiments to date, as
was discussed in section II. Qualitatively however the picture is that at
large coupling the metallic state survives effect of scattering off
impurities due to the reduction in the DOS.

\subsection{Higher order effects. Aronov - Altshuler effect revisited.}

\subsubsection{Higher order corrections to the vertex function and
conductivity}

\begin{figure}[t]
\centerline{ \psfig{file=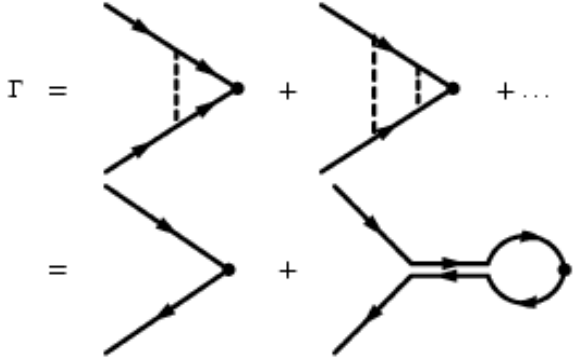,width=0.5\textwidth}}
\caption{Vertex correction due to interactions with disorder (dashed lines)
in the conventional notations \protect\cite{Altshuler} (first line) and in
the present paper notations (second line).}
\label{fig8}
\end{figure}

\begin{figure}[t]
\centerline{ \psfig{file=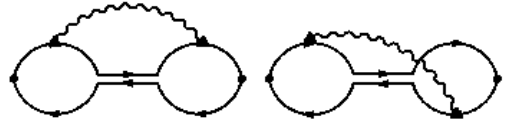,width=0.4\textwidth}}
\caption{Major contributions to the conductivity in perturbation theory}
\label{fig12}
\end{figure}

It was shown in Ref.~\onlinecite{Lee} that in perturbation theory that when
one sums up all the corrections to the vertex part (the $\psi \psi \rho $
condensate where $\rho $ is the density field that couples to static photon)
shown on Fig.~\ref{fig8}, it becomes proportional to the diffusive pole $%
1/(\omega \tau +Dp^{2})$. The same expressions are also shown in Fig.~\ref%
{fig8} in our notations as a sum of leading and the next to leading terms in
the steepest descent expansion. In this picture however the diffuson
propagator is considered in the noninteracting theory. The vertex part
enters high-order diagrams creating logarithmically divergent corrections
which strengthen (in the singlet sector) weak localization. The major
diagrams involving the singular vertex part contributing to conductivity are
given on Fig.~\ref{fig12} (some other contributions cancel, see Ref. %
\onlinecite{Altshuler}). The physical interpretation of this phenomenon is
that electrons scatter coherently on Friedel oscillations due to density
fluctuations. \cite{Aleiner}

\begin{figure}[b]
\centerline{ \psfig{file=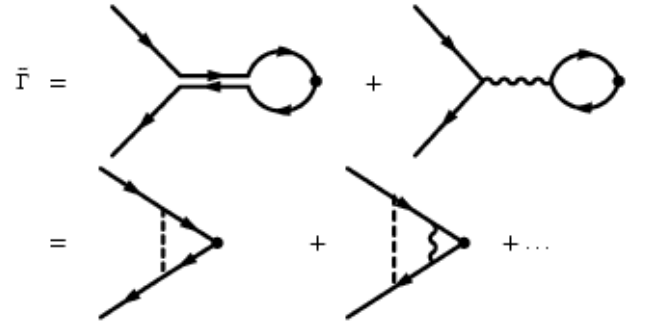,width=0.5\textwidth}}
\caption{Vertex corrections due to mixing of diffusons with static photons
in the present paper notations (first line) and in the conventional
notations \protect\cite{Abrikosov} (second line)}
\label{fig13}
\end{figure}

Without the crucial pole factor the Aronov - Altshuler's corrections do not
diverge in the infrared. In the strong coupling regime considered in this
paper, due to mixing of diffusons with static photons, the vertex is given
in Fig.~\ref{fig13} has much softened small momentum asymptotic: $1/(\omega
\tau +Dp^{2})$. It can be easily seen that this softening is quite enough to
render the contributions like those on Fig.~\ref{fig12} (which is of the
order $1/N_{s}$ namely higher than the weak localization one) finite. One
therefore would ask how this can be understood diagrammatically in terms of
conventional disorder coupling \cite{Abrikosov} ? The point is that the
mixing effectively sums up diagrams to all orders in Coulomb coupling 
$r_{s}$
Fig.~\ref{fig13}. Each one of these is divergent, while their sum is not.
This is quite analogous to disappearance of IR divergencies due to long
range photon "chains" after the RPA diagrams are summed.

\subsubsection{Density of states near Fermi level at large coupling}

The Aronov - Altshuler's corrections to conductivity are directly related to
the downturn cusp in DOS due to Coulomb interaction. In 2D the cusp is given
by \cite{Altshuler}%
\[
\delta N(\varepsilon )\propto \log [\varepsilon ]
\]%
again due to the renormalization of the vertex, see Fig.~\ref{fig8}. It was
claimed that this is precisely what was observed in tunnelling junction
experiments in disordered metal films. \cite{McMillan,Butko} In our
approach, due to Anderson - Higgs mechanism, this renormalization is greatly
reduced. An alternative explanation at very strong coupling and significant
disorder might be the leading order reduction of DOS discussed in section
IID, see Fig.~\ref{fig6}. This does not contradicts the experiments in
metals since in these experiments disorder is large (even very large), while
the coupling $r_{s}$ is quite small. Of course if the density is
sufficiently high the screening can no longer be neglected and the Aronov -
Altshuler's effect becomes dominant. However, as we mentioned before, in
very clean 2DEG samples the disorder can reduce the screening and our
approach of neglecting the screening at the leading order becomes more
appropriate. In this case higher orders will not be large enough to
undermine this assumption. One will get again a reduction in the DOS, but
for entirely different reason. A related issue is emergence of the Coulomb
gap commented on in the next section.

\section{Summary and Discussion}

To summarize we present a consistent gauge invariant approach to disordered
strongly interacting electron gas in 2D. Physically the basic phenomenon is
the reduction in the DOS at Fermi level due to strong Coulomb repulsion.
This in turn suppresses both screening and scattering of impurities
stabilizing the metallic state against weak localization effects. Formally
the approach consists of two steps. The first is variational (or "self
consistent"): the most general quadratic states were considered and one with
minimal energy identified. At this stage the "RPA" screening is neglected
assuming it is sufficiently weakened by disorder so that higher orders are
small. The second step is steepest descent perturbative expansion (which
also can be identified as an expansion in parameter $1/N_{s}$ with $N_{s}$
number of spin components or valleys). Although the general philosophy of
the steepest descent expansion is not drastically different from the one
adopted in other works (see for example Ref.~\onlinecite{Belitz}), two
rather independent observations were made. The first is that the exchange
part of Coulomb interaction leads at strong coupling to a significant
reduction of the DOS near the Fermi surface and to via the reduction
suppresses the disorder effects. The second is that when the steepest
descent expansion procedure, if followed consistently, mixing between static
photons and diffusons not only causes Debye screening of the photon, but
leads in addition to a softening of the diffusion pole. This in turn leads
to a number of observable consequences like a significant modification of
the vertex part and consequently of the Aronov - Altshuler contribution to
conductivity. The contribution becomes regular in the strong coupling, not
logarithmic (the absence of the effect of the DOS reduction due to
interaction in Ref. 17 can be traced to an approximate calculation of the
exchange diagram in their Eq.~(16) ).

In this section we discuss several general questions and assumptions and
relation of our work to other attempts to incorporate the long range Coulomb
interactions into the theory of disordered electron gas.

The theory of disordered electron gas relies to large extent on existence of
massless collective modes, diffusons and Cooperons. It is tempting to
interpret these excitations as Goldstone bosons of some symmetry breaking
(with several complications arising from the quenched disorder, see Ref.~%
\onlinecite{Stone}). The $\sigma $ model approach initiated by Wegner \cite%
{Wegner} and others \cite{Hikami,Pruisken} long ago and developed and
applied to the Coulomb interaction case recently by Baranov et al. \cite%
{Baranov} start from and assumption that the $G=Sp(2N)\otimes Sp(2N)$
symmetry of free disordered electron gas is spontaneously broken down to
diagonal subgroup $H=Sp(2N),$ where $N$ enumerates replica, spin and
Matsubara indices. We have shown in section III however that diffusons mix
with photon and become "harder" than standard Goldstone bosons. This might
signal that the $\sigma $ - model approach should be modified to incorporate
the Anderson - Higgs mechanism. Actually the need for such a modification
can be found in recent remarkable work of Ref.~\onlinecite{Baranov} and we
comment on this now.

Unfortunately the presence of strong Coulomb interactions explicitly breaks
a subgroup of $G$. An example of the explicitly broken symmetry
transformation is $\delta \psi _{pn}=\overline{\psi }_{-p,n},$ $,\delta
\overline{\psi }_{p,n}=-\psi _{-p,-n}$ for positive Matsubara frequencies $%
n>0$ and $\delta \psi _{pn}=\overline{\psi }_{-p,-n}$, $\delta \overline{%
\psi }_{p,n}=-\psi _{-p,-n}$ for negative Matsubara frequencies $n<0.$The
symmetry is broken by both the frequency term $\sum\limits_{p,n}\overline{%
\psi }_{pn}^{a}\left( -i\omega _{n}\right) \psi _{pn}^{a}$ and by the
Coulomb interaction term. However while the breaking by the frequency term
is "soft" and insignificant, as far as static quantities like DC
conductivity are concerned, it was shown \cite{Baranov} that the Coulomb
interaction effectively represented on the $\sigma $ - model level by the
"square of trace" operator\ (Eq.~(2.1) in Ref.~\onlinecite{Baranov}) is
relevant and cannot be reduced to a soft breaking. Baranov et al. notice
that at large distances the diffusion is suppressed which coincides with our
Eq.~(\ref{diffuson}). At short distance scales the electrons are diffusive.
We believe that the Anderson - Higgs mechanism can be treated approximately
within the $\sigma $ - model approach as long as the mixing is small. This
is well known problem in Quantum Field Theory \cite{Kugo} under the name of
"gauged $\sigma $ - models". These issues might become clearer when the
present approach is extended beyond 2D (say to $2+\varepsilon $). Work on
this is in progress. A related issue is understanding the difference between
diffusons and Cooperons. Within the $\sigma $- model approach the $%
Sp(2N)\otimes Sp(2N)$ symmetry forces the Cooperon and the diffuson
propagators to be same. It is precisely an explicit (not spontaneous)
symmetry breaking due to Coulomb interactions that makes the hardening of
the diffuson (mixing with photon), while leaving Cooperon intact possible.

The fact that Coulomb interaction modifies diffusion at large distances was
also discussed in Ref.~\onlinecite{Si} and this might lead also to
suppression of weak localization. The work however was criticized, \cite%
{Sicomment,Baranov,Aleiner} that the vertex parts were not taken into
account or alternatively the treatment is not gauge invariant. Our work
explicitly shows that despite the fact that diffuson is "harder" (although
still massless) at large distances, the Cooperon is not. Therefore although
weak localization is suppressed, the suppression is much weaker and
completely different. The logarithmically divergent contribution to
conductivity includes Cooperon.

It was shown by Efros and Shklovskii \cite{Efros} on basis of a heuristic
argument with \ plausible assumptions about the nature of the localized
electronic states (neglecting their overlaps) that there should be a Coulomb
gap in the strongly interacting electron gas. As in the case of $\sigma $
models with Coulomb interactions, \cite{Baranov} it is not clear from the
first orders in our scheme whether the reduction of the density of state is
along the line of their argument. We believe this is unlikely due to the
fact that they neglect the effect of exchange on the "states" $\psi _{i}$
defined there. It is also not clear at this point whether the opening of
Coulomb gap that Baranov et al. \cite{Baranov} deduce on the basis of the $%
2+\varepsilon $ expansion is related to the reduction in density of states
due to exchange.

One can extend the approach presented here to the "self consistent" scheme
initiated by Vollhardt and developed to include Coulomb interactions by
Sadovsky. \cite{Sadovsky} This will allow quantitative study of the
insulating state and of the Coulomb gap. Note however that the "gap"
equations of Ref.~\onlinecite{Sadovsky} follow the perturbative Aronov -
Altshuler contributions, while the "self consistent" form of our
conductivity contributions Eq.~(\ref{drude}) and Eq.~(\ref{corr}) will
contain different diagrams. The work on this is in progress.

At last we briefly comment on two general assumptions made. The first is the
spatial homogeneity. It is clear \cite{Ceperley} that clean and even
disordered 2DEG \cite{Tanatar} at sufficiently strong coupling become
inhomogeneous Wigner crystals or "glass". It was even speculated \cite%
{Chakravarty} that the Wigner crystallization (which occurs around 
$r_{s}=40$
in clean systems) might be related to the observed metal - insulator
transition. Following general argument can be advanced against such a
scenario. It has been observed recently that in several clean systems of
thermally fluctuating repelling objects the homogeneous state (liquid or
gas) exists down to zero temperature. One such a system is one component
classical plasma. \cite{Rosenfeld} Another is a system of vortex lines in
type II superconductors. \cite{Li} The latter is quite analogous to 2DEG.
The difference is that thermal fluctuations should be replaced by quantum
and bosonic field by fermionic (statistics is quite unimportant in the low
density limit though). To be sure the energy of the solid is lower, so below
the melting point the liquid state is metastable (in conventional liquids
for which in addition to repulsive interaction there is a long range
attractive force, the metastable state ceases to exist at spinodal point).
It is reasonable to assume (and it was demonstrated recently \cite{Li2})
that disorder favors homogeneous state over a structured crystal). Therefore
transition to a Wigner crystal or glass state would occur at much higher
couplings than metal - insulator transition and the relevant state is
homogeneous as was assumed in the present paper.

Another assumption commonly made is that the replica symmetry used to derive
our starting point Eq.~(\ref{Aspin}) was assumed to be unbroken. This means
that we neglected a possibility of "electron glass". \cite{Dobrosavljevic}
This is a distant possibility in the quasimetallic state since, as we argued
in the paper, reduction in the DOS due to long range interactions make
disorder less favored. Eventually in the insulating state glassy behaviors
will eventually prevail.

\begin{acknowledgements}

We are grateful to our colleagues in Hsinchu Profs. T.K Lee, A.
Voskoboynikov, H.H. Lin, S.Y. Hsu, C.Y. Mou and V. Gudmundsson and
especially J.J. Lin for numerous conversations and encouragement.
One of us (B.R.) thanks Profs. S. V. Kravchenko, M. P. Sarachik,
A. Gold and Q. Si for sharing their insight during their visits to
NCTS. He also thanks especially to E. Kogan for important clarification of 
details of
Altshuler - Aronov work and other members of the Theoretical
Physics Group of Bar Ilan University, where part of this work has been done, 
for
encauragement. The work
was supported by the NSC of Taiwan grant No
91-2811-M-009-006 and NCTU-Bar Ilan cooperation grant. One of us (M.-T.) 
also thanks the support of
the Vietnam National Program of Basic Research on Natural Science,
Project No 4.1.2.

\end{acknowledgements}

\appendix*

\section{Approximate solution of the minimization equations}

\subsection{Expansion in small $r_{s}$}

From Eq.~(\ref{simpE}) we observe that at small coupling $r_{s}\ll 1$, $%
\widehat{e}_{\varepsilon }$ starts from the first order: $e_{\varepsilon }=%
\mathrm{sign}[\varepsilon ]\left( r_{s}e_{\varepsilon }^{(1)}+...\right) $.
Substituting this into Eq.~(\ref{simpQ}) immediately gives the leading term
for $q_{\omega }$:
\begin{equation}
\widehat{q}_{\omega }^{(0)}=\mathrm{sign}[\omega ]\frac{1}{4\lambda }=%
\mathrm{sign}[\omega ]\widehat{q}^{(0)}.  \label{q0}
\end{equation}%
Therefore we expand $\widehat{q}_{\omega }=\mathrm{sign}[\omega ]\left(
\widehat{q}^{(0)}+r_{s}\widehat{q}_{\omega }^{(1)}+...\right) $. 
Then the
leading order contribution to $e_{\varepsilon }$ can be computed:%
\begin{eqnarray}
e_{\varepsilon }^{(1)} &=&\frac{\sqrt{2}}{\pi ^{2}}\int_{\omega
>0,\varepsilon ^{\prime }}\kappa \lbrack \varepsilon -\varepsilon ^{\prime 
}]%
\frac{\varepsilon ^{\prime }+e_{\varepsilon ^{\prime }}}{\left( \varepsilon
^{\prime }+e_{\varepsilon ^{\prime }}\right) ^{2}+(\widehat{\omega }+%
\widehat{q}^{(0)})^{2}}  \nonumber \\
&=&\frac{\sqrt{2}}{\pi ^{2}}\int_{\varepsilon ^{\prime }}\kappa \lbrack
\varepsilon -\varepsilon ^{\prime }]\mathrm{sign}(\varepsilon ^{\prime
})\left( \frac{\pi }{2}-\text{Arc}\tan \frac{\widehat{q}^{(0)}}{|\varepsilon
^{\prime }|}\right) .  \nonumber \\
&&  \label{e1}
\end{eqnarray}%
We use here the constant $\widehat{q}_{\omega }$ approximation. Substituting
this into Eq.~(\ref{simpQ}) one obtains:
\begin{eqnarray}
\widehat{q}^{(1)} &=&-\frac{3\sqrt{2}(\widehat{q}^{(0)})^{2}}{2\pi ^{2}}%
\int_{\varepsilon }\frac{\kappa \lbrack \varepsilon ]}{(\varepsilon ^{2}+4%
\widehat{q}^{(0)2})}  \nonumber \\
&=&-\frac{3}{2^{15/2}\pi ^{3}\lambda ^{2}}G_{33}^{22}\left( (4\lambda
)^{2}\Big|_{1,1}^{3/2,3/2}\right) ,  \label{q1}
\end{eqnarray}%
where the Meijer function $G_{pq}^{mn}\left( z\Big|\null%
_{b_{1},...,b_{q}}^{a_{1},...,a_{p}}\right) $ are defined in Ref.~%
\onlinecite{Gradstein}. We observe that it is negative, namely the long
range interaction reduced the effect of disorder! Diagrammatically the
minimization equations sum all the rainbows including both disorder and
interaction, Fig.~2. In perturbation theory one evaluated a diagram with one
photonic rainbow and arbitrary number of disorder lines, namely diagram
Fig.1 with disordered Green's function. Its imaginary part is precisely $%
\widehat{q}^{(1)}$. The actual expansion parameter is $\sqrt{2}r_{s}/\pi
^{2} $ rather than $r_{s}$ as can be seen from comparison of the
perturbative and exact solutions. Therefore perturbation theory breaks down
completely at $r_{s}\sim 10$.

\subsection{Expansion in small $1/\protect\lambda $}

It is important also to obtain an analytical solutions of the minimization
equations at large $r_{s}$. This turns out to be possible for relatively
clean case in which we can expand in $1/\lambda $. The leading order is the
clean solution already discussed in subsection IIB: $e_{\varepsilon
}^{[0]}=e_{\varepsilon }^{[0]}+e_{\varepsilon }^{[1]}/\lambda +...$with $%
e_{\varepsilon }^{[0]}=\sqrt{2}r_{s}\kappa _{1}[\varepsilon ]/\pi $. The
leading term for $\widehat{q}_{\omega }=\widehat{q}_{\omega }^{[1]}/\lambda 
+%
\widehat{q}_{\omega }^{[2]}/\lambda ^{2}+...$is:%
\begin{equation}
\widehat{q}_{\omega }^{[1]}=\frac{1}{4\pi }\int_{\varepsilon 
}\frac{\widehat{%
\omega }}{\left( \varepsilon +e_{\varepsilon }^{[0]}\right) ^{2}+\widehat{%
\omega }^{2}}.  \label{q1D}
\end{equation}%
The correction to dispersion relation can also be computed in a quite
compact form:
\begin{equation}
e_{\varepsilon }^{[1]}=-\frac{r_{s}}{2\sqrt{2}\pi ^{2}}\int_{\varepsilon
^{\prime },\varepsilon ^{\prime \prime }>0}\frac{\kappa \lbrack \varepsilon
-\varepsilon ^{\prime }]-\kappa \lbrack \varepsilon +\varepsilon ^{\prime 
}]%
}{\left( \varepsilon ^{\prime }+e_{\varepsilon ^{\prime }}^{[0]}+\varepsilon
^{\prime \prime }+e_{\varepsilon ^{\prime \prime }}^{[0]}\right) ^{2}}.
\label{e1D}
\end{equation}%
The actual expansion parameter is $1/4\pi \lambda $ rather than $\lambda $
as can be seen from comparison of the perturbative and exact solutions.
Therefore perturbation theory breaks at $\lambda \sim 0.1$. The results are
marked by dotted lines on Fig.~\ref{fig4} and \ref{fig5}. Generally
numerical results agree with both perturbative expansions.

\end{document}